\documentclass[11pt]{article}

\usepackage{graphicx}

\usepackage{amssymb}

\begin{document}
\tolerance=10000
\vskip-1.5truecm
\title{
{\bf Continuous time random walk,} \\ 
{\bf Mittag-Leffler waiting time and}\\  
{\bf fractional diffusion: mathematical aspects}
}
\author{
Rudolf GORENFLO
\\ Dept. of  Mathematics \& Computer Science, Freie Universit\"at  Berlin,
\\  Arnimallee  3, D-14195 Berlin, Germany.
 \\ E-mail: {\tt gorenflo@mi.fu-berlin.de}
\and
Francesco MAINARDI
\\ Department of Physics, University of Bologna, and INFN,
\\ Via Irnerio 46, I-40126 Bologna, Italy.
 \\ E-mail: {\tt francesco.mainardi@unibo.it}
}
\vskip -0.5truecm
 \date{{Revised Version: May 2008}}
\maketitle
\font\sm=cmr10
\font\note=cmr10 at 10 truept
\vskip -0.75 truecm
\noindent
{\sm 
Invited lecture by 
R. Gorenflo  
at the 373 WE-Heraeus-Seminar on  Anomalous Transport: 
Experimental Results and Theoretical Challenges,
Physikzentrum Bad-Honnef (Germany), 12-16 July 2006. 
It will appear in the book {\it Anomalous Transport: Foundations and Applications}
edited by R. Klages, G. Radons and I.M Sokolov, as Chapter 4, pp. 93-127,   
WILEY-VCH, Weinheim, Germany  (2008).}
$\null$                                                                                                                                                                                                                                                    
\vskip -2.0truecm
\begin{abstract}
\noindent
We show the asymptotic long-time equivalence of a generic power law
waiting time distribution to the Mittag-Leffler waiting time distribution,
characteristic for a time fractional
continuous time random walk.
This asymptotic  equivalence is effected by a combination of ``rescaling"
time and ``respeeding" the relevant renewal process followed
by a  passage to a limit for which we need a suitable
relation between the parameters of rescaling and respeeding.
As far as we know such procedure has been first  applied in the
1960s
by Gnedenko and Kovalenko in their
theory of ``thinning" a renewal process.
Turning our attention to spatially one-dimensional continuous time random walks 
with a generic power law jump distribution,
``rescaling" space can be interpreted as a second kind of ``respeeding"
which then, again under a proper relation between the relevant parameters leads
in the limit to the space-time fractional diffusion  equation.
Finally, we treat the `time fractional drift" process as a properly
scaled limit of the counting number of a
Mittag-Leffler renewal process.
\end{abstract}


\tolerance=10000


\hyphenpenalty=2000


\def\pni{\par \noindent}
\def\vsh{\smallskip}
\def\s{\smallskip}
\def\vs{\medskip}
\def\vvs{\bigskip}
\def\vvvs{\bigskip\medskip} 
\def\vsp{\vsh\pni}
\def\vsn{\vsh\pni}
\def\cen{\centerline}
\def\ra{\item{a)\ }} \def\rb{\item{b)\ }}   \def\rc{\item{c)\ }}
\def\eg{{\it e.g.}\ } \def\ie{{\it i.e.}\ }


\def\sg{\hbox{sign}\,}
\def\sgn{\hbox{sign}\,}
\def\sign{\hbox{sign}\,}
\def\e{{\rm e}}
\def\exp{{\rm exp}}
\def\ds{\displaystyle}
\def\dis{\displaystyle}
 \def\q{\quad}    \def\qq{\qquad} 
\def\lan{\langle}\def\ran{\rangle}
\def\lt{\left} \def\rt{\right}  
\def\lra{\Longleftrightarrow}
\def\d{\partial}
\def\dr{\partial r}  \def\dt{\partial t}
\def\dx{\partial x}   \def\dy{\partial y}  \def\dz{\partial z}
\def\rec#1{{1\over{#1}}}
\def\zr{z^{-1}}



\def\hatt{\widehat}
\def\epsilons{{\widetilde \epsilon(s)}}
\def\sigmas{{\widetilde \sigma (s)}}
\def\fs{{\widetilde f(s)}}
\def\Js{{\widetilde J(s)}}
\def\Gs{{\widetilde G(s)}}
\def\Fs{{\wiidetilde F(s)}}
 \def\Ls{{\widetilde L(s)}}
\def\L{{\mathcal L}} 
\def\F{{\mathcal F}} 


\def\NN{{\bf N}}
\def\RR{{\bf R}}
\def\CC{{\bf C}}
\def\ZZ{{\bf Z}} 


\def\I{{\cal I}}  
\def\D{{\cal D}}  


\def\Gc{{\cal {G}}_c}   \def\Gcs{\barr{\Gc}} 
\def\Gs{{\cal {G}}_s}   \def\Gss{\barr{\Gs}} 
\def\Gck{\hatt{\Gc}} 
\def\args{(x/ \sqrt{D})\, s^{1/2}}
\def\argsa{(x/ \sqrt{D})\, s^{\beta}}
\def\argx{ x^2/ (4\,D\, t)}


\def\erf{\hbox{erf}}  \def\erfc{\hbox{erfc}}


\def\uks{{\widehat{\widetilde {u}}} (\kappa,s)}

\def\psikappa{\psi_\alpha^\theta(\kappa)}


\font\bfs=cmbx12 scaled\magstep1


\newpage

\section{Introduction}

The purpose  of this paper is to outline the fundamental role the 
Mittag-Leffler function  in renewal processes that are relevant in the 
theories of  anomalous diffusion. 
As a matter of fact the interest in this function in statistical physics and 
probability theory has recently  increased as is shown  by the large 
number of papers published  since 1990 of which 
a brief (incomplete)  bibliography
includes 
\cite{Barkai-Silbey_JPCB00,Barkai-Sokolov_PhysicaA07,GAR_Vietnam04,
Gorenflo-Mainardi_INDIA03,
Gorenflo-Mainardi_CARRY04,Hilfer_FRACTALS95,Hilfer_PhysicaA03,
Hilfer-Anton_PRE95,
Huillet_JPhysA02,Huillet_FRACTALS02,Kozubowski-Rachev_99,
Mainardi_FCAA05,Mainardi_PhysicaA00,Metzler-Barkai-Klafter_PRL99,
Pillai_90,Scalas_PRE04,Sokolov-Klafter-Blumen_PRE01,Weron-Kotulski_PhysicaA95}.
\vsp    
 In this paper we develop a theory for long-time
behaviour of a renewal process with a generic power law waiting distribution
of order $\beta $, $0<\beta \le 1$
(thereby for easy readability  dispensing with decoration by
a slowly varying function).
To bring the distant future into near sight
we   change the unit of time from $1$ to $1/\tau $,
$0 <\tau  \ll 1$.
\vsp
For the random waiting times $T$ this means replacing
$T$ by $\tau T$.
Then, having very many events in a moderate span of time
we compensate this compression by respeeding the whole process,
actually slowing it down so that again we have a moderate
number of events in a moderate span of time.
We will relate the rescaling factor $\tau $ and the respeeding factor $a$
 in such a way that in the limit $\tau \to 0$ we have a reasonable
process, namely one whose waiting time distribution
is the Mittag-Leffler waiting time distribution whose density
is
$$ \phi^{ML} (t) =-  \frac{d}{dt} E_\beta (-t^\beta )\,,\quad
0<\beta \le 1\,, \eqno(1.1)$$
with the Mittag-Leffler function
$$ E_\beta (z) := \sum_{n=0}^{\infty} \frac{z^n}{\Gamma (\beta n+1)}\,,
\quad z\in \CC\,, \, \quad \beta >0\,.
\eqno(1.2)$$
We will call the renewal process with waiting time density 
$ \phi^{ML} (t)$ the {\it Mittag-Leffler (renewal) process}.
This process can be seen as a fractional generalization of the Poisson process, see
\cite{Mainardi_VIETNAM04}.
\vsp
Our method is, in some sense, analogous to the one applied
 in the
Sixties of the past century by Gnedenko and Kovalenko
\cite{GnedenkoKovalenko_QUEUEING68} in their
analysis  of {\it thinning} (or {\it rarefaction}) of a renewal process.
They found, under certain power law assumptions, in the infinite  thinning limit,
for the waiting time density the Laplace transform
$1/(1+s^\beta)$ but did not identify
it as a Mittag-Leffler type function.
In Section 2, we provide, in our notation, an outline of the thinning theory
for renewal processes essentially following Gnedenko and Kovalenko.
Their method has inspired us for the reatment of our problems.
\vsp
As we consider our renewal process formally as a {\it continuous time random walk}
(CTRW) with constant non-random  jumps 1 in space (for the counting function $N(t)$,
in Section 3 we embed ab initio our theory into that of the CTRW,
thus being in the position to treat the theory of a time fractional
CTRW as limiting case of a CTRW with power law waiting time distribution.
In this context the pioneering paper by Balakrishnan
\cite{Balakrishnan_85} of 1985 deserves to be mentioned.
Balakrishnan   already found the importance of the Laplace transform
$1/(1+s^\beta )$ in the time fractional CTRW and
time fractional diffusion, but also did not identify
it as the Laplace transform of $ \phi^{ML} (t)$.
Then, in 1995 Hilfer and Anton \cite{Hilfer-Anton_PRE95}, see also
\cite{Hilfer_FRACTALS95,Hilfer_PhysicaA03}, showed
that this waiting time density is characteristic for the time fractional CTRW
and can be  expressed in terms
of the Mittag-Leffler function in two parameters, that is
$$ \phi^{ML} (t) = t^{\beta-1}\, E_{\beta,\beta} (-t^\beta )\,,\quad
0<\beta \le 1\,, \eqno(1.3)$$
with the generalized Mittag-Leffler function 
$$ E_{\beta,\gamma} (z) :=
\sum_{n=0}^{\infty} \frac{z^n}{\Gamma (\beta n+\gamma)}\,,
\quad z\in \CC\,, \quad \beta>0\,, \quad \gamma \in \RR\,.
\eqno(1.4)$$
The form (1.3) is equivalent to the form (1.1)
that we prefer as it exhibits visibly also the cumulative probability function,
the {\it survival function}, $E_\beta(-t^\beta)$.
\vsp
We explain in Section 4 two manipulations, {\it rescaling} and {\it respeeding}
and use these in Section 5 to deduce the asymptotic universality of 
the Mittag-Leffler waiting time density under a power law assumption for the
original waiting time.
Then, in  Section 6,  assuming a suitable power law also for the spatial jumps
we show that by a rescaling of the jump widths by a positive factor $h$
(that means a change of the unit of space from 1 to $1/h$ to bring
into near sight the far-away space) another respeeding is effected,
now an acceleration, that in the limit $h \to 0$
(under a proper relation between $h$ and $\tau $) leads
to space-time fractional diffusion.
\vsp
In Section 7, we pass to a properly scaled limit for the counting function
$N(t)$ of a renewal process (again under power law assumption)
and obtain the time fractional drift process
(viewing $N(t)$  as a spatial variable).
\vsp
We  will extensively work with the transforms of Laplace and Fourier,
so easing calculations and proofs of {\it convergence in distribution}
(also called ``weak convergence") for our passages to the limit.
\vsp
Essentially, we treat in this paper three topics. First, in Section 2, the thinning of 
a pure renewal process. Second, in Sections 3-6, under power law assumption for the waiting time,
the asymptotic relevance of the Mittag-Leffler law, and then the general CTRW
with special attention to space and time transition limits to fractional diffusion.
As a third  topic, in Section 7, we investigate the long time behaviour 
of the Mittag-Leffler renewal process.
\vsp
Essential properties of the derivative of fractional order in time
and in space are given in Appendix A and Appendix B, respectively.
Finally, in  Appendix C we give details on the two special functions
of the Mittag-Leffler type that play a fundamental role in this paper,
the Mittag-Leffler survival probability and the Mittag-Leffler
waiting time density. 

\section{An outline of the Gnedenko-Kovalenko  theory of thinning}   

The {\it thinning} theory for a renewal process has been considered
in detail by  Gnedenko and Kovalenko \cite{GnedenkoKovalenko_QUEUEING68}.
We must note that other authors, like Sz\'antai
\cite{Szantai_71a,Szantai_71b} speak of {\it rarefaction}
in place of thinning.
Let us sketch here the essentials of this theory:
in the interest of transparency and easy
readability we avoid the possible decoration
of the relevant power law by multiplying it with a 
{\it slowly varying function}.
As usual  we call a (measurable) positive function $a(y)$ 
{\it slowly varying at zero} if
$a(cy)/a(y) \to 1$ with $y \to 0^+$ for every $c>0$,
{\it slowly varying at infinity}
 if $a(cy)/a(y) \to 1$ with $y \to +\infty$ for every $c>0$.
A standard example of a slowly varying function at zero and at infinity is 
$|\log y|^\gamma$, with $\gamma \in \RR$.
\vsp
Denoting by $t_n$,\, $n=1,2,3, \dots$
the time instants of events of a renewal process, assuming
$0=t_0<t_1<t_2<t_3<\dots $,
with $i.i.d.$  waiting times $T_1 = t_1\,,\,T_k = t_{k}-t_{k-1}$ for $k\ge 2$,
(generically denoted by T),
{\it thinning} (or {\it rarefaction})
means that for each positive
index  $k$  a decision is made:
the event happening in the instant $t_k$ is deleted with probability $p$
or it is maintained with probability $q=1-p$,  $0<q<1$.
This procedure produces a {\it thinned} or {\it rarefied} renewal process
with fewer events (very few events
if $q$ is near zero, the case of particular interest)
in a moderate span of time.
\vsp
To compensate for this loss  we change the unit of time
so that we still have not very few  but still a moderate number
of events in a moderate span of time.
Such change of the unit of time is
equivalent to  rescaling the waiting time,
multiplying it with a positive factor $\tau $ so that we have
waiting times $\tau T_1,\tau T_2, \tau T_3, \dots$, and
instants $\tau t_1,\tau t_2, \tau t_3,\dots$, in the
rescaled  process.
Our intention is, vaguely speaking, to dispose
on $\tau $ in relation to the rarefaction parameter $q$
in such a way that for $q$ near zero in some sense
the ``average" number of events per unit of time
remains unchanged. In an asymptotic sense
we will make these considerations precise.
\vsp
Denoting by $F(t) = P(T\le t)$ the probability distribution function
of the (original) waiting time $T$,   by $f(t)$ its density
($f(t)$ is a generalized function generating a probability measure)
so that
$F(t) = \int_0^t f(t') \, dt'$,  and analogously by
$F_k(t)$  and $f_k$(t) the distribution 
and density, 
respectively,  of the sum of $k$  waiting times, we have recursively
$$ f_1(t) = f(t) \,,\q
      f_k(t) = \int_0^t  f_{k-1}(t-t') \, dF(t') \,,\;
\hbox{for} \; k \ge 2\,. \eqno(2.1)$$
Observing that after a maintained event the next one of the
original process is kept with probability $q$ but dropped in favour
of the second-next with probability $p\,q$
and, generally, $n-1$ events are dropped  in favour of the
$n$-th-next with probability $p^{n-1}\,q$,
we get for   the waiting time density of the thinned process
the formula
$$ g_q(t) = \sum_{n=1}^\infty q\, p^{n-1}\, f_n(t)\,.\eqno(2.2)$$
With the modified waiting time $\tau \,T$ we have
$$ P(\tau T\le t) = P(T\le t/\tau ) = F(t/\tau )\,, $$
hence the density $f(t/\tau )/\tau $, and analogously for the
density of the sum of $n$ waiting times
$f_n(t/\tau )/\tau $.
The density of the waiting time of the rescaled (and thinned) process
now turns out as
$$ g_{q,\tau} (t) = \sum_{n=1}^\infty q\, p^{n-1}\, f_n(t/\tau)/\tau \,.
\eqno(2.3)$$
\vsp
In the Laplace domain we have
 $\widetilde f_n(s) = \left(\widetilde f(s)\right)^n\,,$
hence (using $p =1-q$)
$$ \widetilde g_q (s)=
\sum_{n=1}^\infty q\, p^{n-1}\,\left(\widetilde f(s)\right)^n
 = \frac{ q\, \widetilde f(s)}{ 1 - (1-q)\, \widetilde f(s)}
 \,,\eqno(2.4)$$
from which by Laplace inversion we can, in principle, construct
the waiting time density of the thinned process.
By  rescaling we get
$$
\widetilde g_{q,\tau}(s)=
\sum_{n=1}^\infty q\, p^{n-1}\,\left(\widetilde f(\tau s)\right)^n
 = \frac{ q\, \widetilde f(\tau s)}{ 1 - (1-q)\, \widetilde f(\tau s)}
 \,.\eqno(2.5)$$
\vs
\noindent
     Being interested in stronger and  stronger
thinning ({\it infinite thinning})
let us now  consider 
 a scale of processes with  the parameters $\tau  $ (of {\it rescaling})
and $q$ (of {\it thinning}), with $q$ tending
to zero  {\it under a scaling relation $q = q(\tau) $
  yet to be specified}.
\vsp
We have essentially two cases for the waiting time distribution:
its expectation value is finite or infinite.
In the first case we put
$$ \lambda  = \int_0^\infty t'\,  f(t')\, dt' < \infty \,.
\eqno (2.6a)$$
In the second case we assume a queue of power law type
(dispensing with a possible decoration by 
a function slowly varying  at infinity)
 $$ \Psi(t) :=  \int_t^\infty f(t')\, dt'
\sim \frac{c}{\beta } t^{-\beta }\,, \; t\to \infty \quad
\hbox{if}\quad 0<\beta <1\,,\eqno(2.6b)$$
Then, by the Karamata theory (see \cite{Feller_71,Widder_BOOK46})
the above  conditions mean in the Laplace domain
$$\widetilde f(s) =1-\lambda  \, s^\beta  + o\left( s^\beta \right)\,,
 \q \hbox{for} \q s \to 0^+\,,\eqno(2.7)$$
with  a positive coefficient $\lambda $ and $0<\beta \le 1$.
The case $\beta =1$  obviously corresponds to the situation with finite
first moment (2.6a), whereas the case $0<\beta <1$ is related
to a power law   queue with
$c= \lambda \,\Gamma(\beta +1)\,\sin(\beta \pi)/\pi\,.$
\vsp
Now, passing  to the limit of $q \to 0$ of infinite thinning under the scaling relation
$$  q = \lambda \, \tau ^\beta \,, \q 0<\beta \le 1\,, \eqno(2.8)$$
between the positive parameters $q$ and $\tau $,
the Laplace transform of the rescaled density
$\widetilde {g_{q,\tau }}(s)$ in (2.5) of the thinned process tends for fixed $s$ to
$$ \widetilde g(s) = \frac{1}{1+s^\beta}\,,  \eqno(2.9)$$
which corresponds to the Mittag-Leffler density
$$ g(t) = - \frac{d}{dt} E_\beta (-t^\beta )
= \phi^{ML}(t)
\,. \eqno(2.10)$$
Let us remark that Gnedenko and Kovalenko obtained
(2.9) as the Laplace transform of the limiting density
but did not identify it as the Laplace transform of a
Mittag-Leffler type function.
Observe that in the special case $\lambda < \infty$ 
we have $\beta=1$,
hence as the limiting process the Poisson process, as formerly shown
in 1956 by R\'enyi \cite{Renyi_56}.  
\vsp


\section{The continuous time random walk (CTRW)}  

The name    {\it continuous time random walk} (CTRW)  became
popular in physics after Montroll, Weiss and Scher (just to cite  the pioneers)
in the 1960's and 1970's published a celebrated series
of papers on random walks for modelling
diffusion processes on lattices, see \eg
\cite{MontrollScher_73,MontrollWeiss_65}, and
the book by Weiss \cite{Weiss_BOOK94} with  references therein.
CTRWs are rather good and general phenomenological models for diffusion,
including processes of anomalous transport,
that can be understood  in the framework of
the classical renewal theory, as stated
\eg in  the booklet by Cox  \cite{Cox_RENEWAL67}.
In fact a CTRW can be considered
as a compound renewal process (a simple renewal process with reward) or
 a  random walk {\it subordinated}
to a simple  renewal process.
\vsp
A spatially one-dimensional CTRW  
is generated by a sequence
of  independent identically  distributed ($iid$)
 positive  random  waiting times $T_1, T_2, T_3, \dots ,$
each having the same probability density function
 $\phi(t)\,,$   $\, t>0\,, $ and
a sequence of $iid$ random jumps $X_1, X_2, X_3, \dots, $
in $\RR\,,$ each having the same probability density
$w(x)\,,$ $\, x\in \RR\,.$
\vsp
Let us remark that, for ease of language, we use
the word density also for generalized functions
in the sense of Gel'fand \& Shilov \cite{GelfandShilov_64},
that can be interpreted as probability measures.
Usually the {\it probability density functions} are abbreviated
by  $pdf$.
We recall that $\phi (t) \ge 0$ with $\int_0^\infty \phi (t)\, dt =1$
and  $w(x)  \ge 0$ with $\int_{-\infty}^{+\infty} w (x)\, dx =1$.
\vsp
Setting
$t_0=0\,,$ $\, t_n = T_1+T_2 + \dots T_n$ for $n \in \NN\,,$
the wandering particle 
makes a jump
of length $X_n$ in instant $t_n$,
so that its position is $x_0=0$ for
$0\le t <T_1= t_1\,,$       and
$x_n =    X_1 + X_2 + \dots X_n\,,$
for $ t_n \le t < t_{n+1}\,. $
We require the distribution of
the waiting times
and  that of the jumps to be independent
of each other.
So, we have a compound renewal process (a renewal process with
       reward), compare  \cite{Cox_RENEWAL67}.
\vsp
By natural probabilistic arguments we arrive at the
{\it integral  equation} for the probability density   $p(x,t)$
(a density with respect to the variable $x$)
of the particle being in point $x$ at instant $t\,, $
see  \eg
\cite{Gorenflo-Mainardi_INDIA03,Gorenflo_KONSTANZ01,Mainardi_PhysicaA00,Scalas_PhysA05,SGM_00,Scalas_PRE04},
$$
   p(x,t) =  \delta (x)\, \Psi(t)\, +
  \int_0^t  \!\!  \phi(t-t') \, \lt[
 \int_{-\infty}^{+\infty}\!\!  w(x-x')\, p(x',t')\, dx'\rt]\,dt'\,,
\eqno(3.1)  $$
 in which  the {\it survival function}
 $$\Psi(t) = \int_t^\infty \phi(t') \, dt' \eqno(3.2)$$
denotes the probability that at instant $t$ the particle
 is still sitting in its starting position
$x=0\,. $
Clearly, (3.1) satisfies the initial condition
$p(x,0^+) = \delta (x)$.
\vsp
Note that the {\it special choice}
$$w(x) = \delta (x-1) \eqno(3.3)$$
   gives the {\it pure renewal process}, with
position  $x(t)=N(t)$, denoting the {\it counting function},
and with jumps all of length 1 in positive
direction happening at the renewal instants.
\vsp
For many purposes the integral equation (3.1) of CTRW
can be easily treated
by using the Laplace and Fourier transforms.
Writing these as
$$     {\L} \lt\{ f(t);s\rt\}=  \widetilde f(s)
 := \int_0^{\infty} \!\!\e^{\ds \, -st}\, f(t)\, dt \,, $$
$$ {\F} \lt\{g(x);\kappa\rt\}=  \widehat g(\kappa)
  := \int_{-\infty}^{+\infty} \!\! \e^{\,\ds +i\kappa x}\,g(x)\, dx
     \,,
$$
then in the Laplace-Fourier domain
Eq. (3.1)  reads
$$ \widehat{\widetilde p}(\kappa ,s)
 =   {1-\widetilde\phi(s)  \over s} +
 \widetilde \phi(s) \,  \widehat w(\kappa )\,
   \widehat{\widetilde p}(\kappa ,s) \,, \eqno(3.4)$$
Introducing formally in the Laplace domain the auxiliary function 
$$ \widetilde{H} (s) =
\frac{1- \widetilde{\phi}(s) }{ s\, \widetilde{\phi}(s)}
= \frac{\widetilde{\Psi}(s) }{\widetilde{\phi}(s)}
\,, \q \hbox{hence}\q
   \widetilde{\phi}(s) =  \frac{1}{1+s \widetilde{H} (s)} \,,
 \eqno(3.5)$$   
 and assuming that its Laplace inverse $H(t)$ exists, we get,
 following \cite{Mainardi_PhysicaA00}, 
in the Laplace-Fourier domain the  equation 
$$  \widetilde{H} (s) \, \left[
  s\widehat{\widetilde p}(\kappa ,s)-1\right] =
  \left[ \widehat w(\kappa )-1\right]\,
   \widehat{\widetilde p}(\kappa ,s)
\,, \eqno(3.6)$$
and in the space-time  domain  the generalized Kolmogorov-Feller equation
$$ \int_0^t   H(t-t')\,
 \frac{\d}{\d t'} p(x,t')\, dt'  =
      -p(x,t) + \int_{-\infty}^{+\infty} w (x-x')\,p(x',t)\,dx',
\eqno(3.7)$$
with $p(x,0) =\delta (x)$. 
\vsp
If  the Laplace inverse $H(t)$ of the formally introduced function $\widetilde H(s)$ does not exist, 
we can formally set
$\widetilde K(s) = 1/\widetilde H(s)$ and multiply  (3.6) with $\widetilde K(s)$.
Then, if $K(t)$ exists, we get in place of (3.7) the alternative
form of the generalized Kolmogorov-Feller equation
$$ 
 \frac{\d}{\d t} p(x,t)\  =
     \int_0^t K(t-t')\,\left[ -p(x,t') + \int_{-\infty}^{+\infty} w (x-x')\,p(x',t')\,dx'
	 \right]\, dt'\,,
\eqno(3.7')$$
with $p(x,0) =\delta (x)$.
\vsp
Special choices of the memory function $H(t)$ are
$  {\mathbf{(i)}}$ and ${\mathbf{(ii)}}$, see eqs (3.8) and (3.12):
\vsp
$$  {\mathbf{(i)}}  \quad H(t)   =  \delta (t)
\q \hbox{corresponding to} \q \widetilde H(s)= 1\,, \eqno(3.8)$$
 giving   the {\it exponential waiting time} with
$$ \widetilde \phi (s) = {\ds \rec{1+s}}, \quad
 \phi (t) =   -  \frac{d}{dt}  \e^{\ds \, -t} =\e^{\ds \, -t}\,, 
 \quad
  \Psi(t) = \e^{\ds \, -t}\,.\eqno(3.9)  $$
In this case we obtain in the Fourier- Laplace domain
$$ 
  s\widehat{\widetilde p}(\kappa ,s)-1 =
  \lt[ \widehat w(\kappa )-1\right]\,
   \widehat{\widetilde p}(\kappa ,s)
\,, \eqno(3.10)$$
and in the space-time domain 
the {\it classical Kolmogorov-Feller equation}
$$ \frac{\d}{ \d t} p(x,t) = - p(x,t)  +
   \int_{-\infty}^{+\infty}   w  (x-x')\,p(x',t)\,dx'\,,
     \quad p(x,0) = \delta(x)\,.
\eqno(3.11) $$
\vsp
$$   {\mathbf{(ii)}}\quad H(t)   = {\ds \frac{t^{-\beta}}{ \Gamma(1-\beta)}}\,,\; 0<\beta <1\,,
\; \hbox{corresponding to} \; \widetilde H(s)= s^{\beta-1}\,,
\eqno(3.12)  $$
giving  the {\it Mittag-Leffler waiting time}
with
$$ \widetilde \phi (s) = {\ds \rec{1+s^\beta }}, \quad
 \phi (t) =       - {\ds \frac{d}{dt} E_\beta (-t^\beta )}
 = \phi^{ML}(t),
 \quad
 \Psi(t) =E_\beta (-t^\beta)\,.\eqno(3.13)  $$
In this case we obtain in the Fourier-Laplace domain
$$ s^{\beta-1} \, \left[
  s\widehat{\widetilde p}(\kappa ,s)-1\rt] =
  \lt[ \widehat w(\kappa )-1\right]\,
   \widehat{\widetilde p}(\kappa ,s)
\,, \eqno(3.14)$$
and in the space-time domain the {\it time fractional Kolmogorov-Feller equation} 
$$   \, _tD_*^\beta \,  p(x,t) =
     -  p(x,t) +   \int_{-\infty}^{+\infty} w(x-x')\,
   p(x',t) \, dx'\,, \quad p(x,0^+) = \delta(x)\,,   \eqno(3.15)$$
   where $\, _tD_*^\beta \,$ denotes the fractional derivative of of order $\beta$
   in the Caputo sense, see Appendix A.  
\vsp
The time fractional Kolmogorov-Feller equation can be also expressed via
the Riemann-Liouville fractional derivative $\,_tD^{1-\beta}$, see again
Appendix A, 
that is
$$ \frac{\d}{\dt}p(x,t) = \,_tD^{1-\beta}\,\left[
 -  p(x,t) +   \int_{-\infty}^{+\infty} w(x-x')\,
   p(x',t) \, dx'\right],     \eqno(3.16)$$
   with  $p(x,0^+) = \delta (x)$.
 The equivalence of the two forms (3.15) and (3.16) is easily proved
 in the  Fourier-Laplace domain by multiplying both sides of Eq. (3.14)
 with the factor $s^{1-\beta}$.   
\vsp
We  note that the choice ${\mathbf{(i)}}$  may be considered as a limit
of the choice ${\mathbf{(ii)}}$ as $\beta =1$.
In fact, in this limit we find   
$\widetilde H(s)  \equiv 1$ so 
$H(t)= t^{-1}/\Gamma(0)\equiv \delta(t)$
(according to a formal representation of the Dirac generalized function \cite{GelfandShilov_64}),
so that Eqs. (3.6)-(3.7) reduce to (3.10)-(3.11), respectively.
In this case the order of the Caputo  derivative
reduces to 1 and that of the R-L derivative to 0, 
whereas the Mittag-Leffler waiting time  law reduces to the
exponential.
\vsp
In the sequel we will formally unite the choices ({\bf i}) and ({\bf ii}) by defining what we call
the Mittag-Leffler memory function
$$  H^{ML}(t)= 
\cases{
{\ds \frac{t^{-\beta}} {\Gamma(1-\beta)}} \,, &  $\hbox{if} \q 0<\beta<1\,,$\cr
{\ds \delta(t)}\,, & $\hbox{if} \q \beta=1 \,,$}
\eqno(3.17)
$$
whose Laplace transform is
$$\widetilde H^{ML}(s)= s^{\beta -1}\,, \q 0<\beta \le 1\,.\eqno(3.18)
$$
Thus we will consider the whole range $0<\beta \le 1$
 by  extending  the Mittag-Leffler waiting time law
in (3.13) to include  the  exponential law (3.9).  
\vsp
{\bf Remark:} Equation (3.7) clearly may be supplemented by an arbitrary
initial probability density $p(x,0)= f(x)$.
The corresponding  replacement of $\delta(x))$
by $f(x)$ in (3.1) then requires in (3.4) multiplication of the term
$(1-\widetilde\phi(s))/s$ by $\widehat f(\kappa)$ and in (3.6) replacement of the
LHS by 
$\widetilde{H} (s) \, \left[
  s\widehat{\widetilde p}(\kappa ,s)-\widehat f(\kappa)\right]$. 
With $p(x,0) =\delta(x)$ we obtain in $p(x,t)$ the fundamental solution of (3.7)

\section{Manipulations: rescaling and respeeding}  

We now consider two types of manipulations on the CTRW
by acting  on its  governing equation (3.7) in its Laplace-Fourier representation (3.6).
\\
{\bf (A): rescaling the waiting time}, hence the whole time axis;
\\
{\bf (B): respeeding the process}.
\vsp
{\bf (A)} means change of the unit of time (measurement).
We replace the random waiting
time  $T$  by a waiting time $\tau T$,
with the positive {\it rescaling factor} $\tau$.
Our idea is to take $0<\tau \ll 1$   in order to bring into near
sight the distant future.
In a moderate span of time we will so have a large number of jump
events. For $\tau >0$  we get the rescaled waiting time density
$$\phi _\tau (t) = \phi (t/\tau )/\tau\,, \q \hbox{hence} \q
\widetilde\phi _\tau (s) = \widetilde\phi (\tau s)\,.  \eqno(4.1)$$
By decorating also the density $p$ with an index $\tau $   we
obtain the rescaled integral equation of the CTRW
in the Laplace-Fourier  domain as
$$ \widetilde{H}_\tau  (s) \, \lt[
  s\widehat{\widetilde {p}}_\tau(\kappa ,s)-1\rt] =
  \lt[ \widehat w(\kappa )-1\rt]\,
   \widehat{\widetilde {p}}_\tau (\kappa ,s)\,,
\eqno(4.2)$$
where, in analogy to (3.5),
$$ \widetilde{H}_\tau (s) =
\frac{1- \widetilde{\phi}(\tau s) }{ s\, \widetilde{\phi}(\tau s)}\,.
\eqno(4.3)$$
\vsp
{\bf (B)}  means multiplying the quantity representing ${\ds \frac{\d}{\dt} p (x,t)}$
by a factor $1/a$, where $a>0$ is the {\it respeeding factor}:
$a>1$ means {\it acceleration}, $0<a<1$  means {\it deceleration}.
In the Laplace-Fourier representation this means multiplying the RHS of Eq. (3.6) by the factor $a$  
since the expression
 $\lt[s\widehat{\widetilde {p}}(\kappa ,s)-1\rt]$
    corresponds to
$ {\ds \frac{\d}{\dt} p (x,t)}$.
\vsp
We now chose
to consider the procedures of rescaling and respeeding
in their combination so that the equation in the transformed domain 
of the  rescaled and respeeded process has the form
$$  \widetilde{H}_\tau  (s) \, \lt[
  s\widehat{\widetilde {p}}_{\tau,a}(\kappa ,s)-1\rt] =
 a\, \lt[ \widehat w(\kappa )-1\rt]\,
   \widehat{\widetilde {p}}_{\tau,a} (\kappa ,s)\,, 
   \eqno(4.4)$$
Clearly, the two manipulations  can be discussed separately:
the choice $\{\tau >0,\, a=1\}$ means {\it pure rescaling},
the choice $\{\tau =1,\, a >0\}$ means {\it pure respeeding}
of the original process.
In the special case  $\tau =1$   we only respeed the
original system; if  $0 <\tau \ll 1$
  we can counteract the compression effected by rescaling to again obtain a
moderate number of events in a moderate span of time by respeeding
(decelerating) with  $0<a \ll 1$.
 These vague notions will become clear as soon as we consider power law
waiting times.
\vsp
Defining now
$$ \widetilde{H}_{\tau,a}  (s) := \frac{\widetilde{H}_{\tau}  (s)}{a}=
\frac{1- \widetilde{\phi}(\tau s) }{ as\, \widetilde{\phi}(\tau s)}\,.
\eqno(4.5)$$
we finally get, in analogy to (3.6), the equation
$$ \widetilde{H}_{\tau,a}  (s) \, \lt[
  s\widehat{\widetilde {p}}_{\tau,a}(\kappa ,s)-1\rt] =
  \lt[ \widehat w(\kappa )-1\rt]\,
   \widehat{\widetilde {p}}_{\tau,a}(\kappa,s)\,.
   \eqno(4.6)
$$
What is the combined effect of rescaling and respeeding on the waiting
time density?
\vsp
In analogy to (3.5) and taking account of (4.5)  we find
$$   \widetilde{\phi}_{\tau ,a}(s) =
        \frac{1}{1+s \widetilde{H}_{\tau ,a} (s)}=
        \frac{1}
{1+s {\ds\frac{1-\widetilde{\phi}(\tau s)}{as \,\widetilde{\phi}(\tau s)}}}
\,,\eqno(4.7)$$
and so, for the deformation of the waiting time density, the
{\it essential formula}
$$   \widetilde{\phi}_{\tau ,a}(s) =
     \frac{a\,\widetilde{\phi}(\tau s) }
     {1- (1-a)\widetilde{\phi}(\tau s)}
\,. \eqno(4.8) $$
\vsp
{\bf Remark}:
The formula (4.8) has the same structure as the thinning
formula (2.5) 
by  identification of  $a$  with $q$. 
In both problems  we have a rescaled process defined by a time scale $\tau$,
and  we send the  relevant factors $\tau$, $a$ and $q$ to zero
under a proper relationship.
However in the thinning theory the relevant independent parameter going to 0 
is that of thinning (actually respeeding) whereas in the present problem 
it is the rescaling parameter $\tau$.


\section{Power laws and asymptotic universality of the Mittag-Leffler waiting
time density}    

We have essentially two different situations for the waiting time distribution
according to its first moment (the expectation value)  being  finite or infinite.
In other words we assume for the waiting time $pdf$ $\phi(t)$
 either
$$ \rho  := \int_0^\infty t'\,  \phi(t')\, dt' < \infty \,, \quad
\hbox{labelled as}\; \beta=1\,,  
\eqno (5.1)$$
or
$$  
\phi(t) \sim c \, t^{-(\beta +1)} \; \hbox{for}\; t \to \infty \quad \hbox{hence} \;
 \; \Psi(t) \sim \frac{c}{\beta} \, t^{-\beta } \,,
 \; 0<\beta<1\,,\; c>0\,. \eqno(5.2)$$
 For convenience we have dispensed in (5.2) with decorating by a slowly varying function at 
 infinity  the asymptotic power law.
Then, by the standard Tauberian theory (see \cite{Feller_71,Widder_BOOK46})
the above  conditions (5.1)-(5.2) mean in the Laplace domain
 the (comprehensive) asymptotic  
form
$$  \widetilde{\phi}(s) =1 -\lambda s^\beta  + o(s^\beta )
\q \hbox{for} \q s \to 0^+\,,
\q  
0<\beta \le 1\,,
\eqno(5.3)$$
where we have 
$$  \lambda = \rho\,,\q \hbox{if}\q \beta=1\,; \; 
 \lambda = c \Gamma(-\beta )=
\frac{c}{\Gamma (\beta+1)} \, \frac{\pi} {\sin(\beta \pi)}\,,
\; \hbox{if}\; 0<\beta<1\,.
 \eqno(5.4) $$
Then, {\it fixing $s$}    as required by the continuity theorem of
probability theory
for Laplace transforms, taking 
$$a =\lambda \tau ^\beta\,,\eqno(5.5)$$
 and {\it sending $\tau $  to zero}, we obtain in the limit
the Mittag-Leffler waiting time law.
In fact,  Eqs. (4.8) and (5.3) imply  as $\tau \to 0$
with  $0<\beta \le 1$,
$$    \widetilde{\phi}_{\tau ,\lambda \tau ^\beta} (s) 
=  \frac{\lambda \tau^\beta\, \left[1-\lambda \tau^\beta s^\beta + o(\tau^\beta s^\beta)\right]}
{1 - (1-\lambda \tau^\beta)\, \left[1-\lambda \tau^\beta s^\beta + o(\tau^\beta s^\beta)\right] }
  \to
\frac{1}{1 + s^\beta }\,, \eqno(5.6)$$
the Laplace transform of $\phi^{ML}(t)$, see (1.1) and Appendix C.
This formula expresses {\bf the asymptotic universality of
the Mittag-Leffler waiting time law} that includes the exponential law for $\beta=1$.
It can easily be generalized to the case of power laws decorated with
slowly varying functions, thereby using the Tauberian theory by Karamata 
(see again \cite{Feller_71,Widder_BOOK46}).
\vsp
{\bf Comment:} The formula (5.6) says that our general power law waiting time
density is
gradually deformed into the Mittag-Leffler waiting time density as
$\tau $ tends to zero.
\vsp
{\bf Remark:} Let us stress here the distinguished character of the Mittag-Leffler
waiting time density ${\ds \phi^{ML}(t)= - \frac{d}{dt} E_\beta(-t^\beta)}$
defined in (1.1).
Considering its Laplace transform   
$$   \widetilde{\phi}^{ML}(s) = \frac{1}{1+s^\beta }\,, \quad  0<\beta \le 1\,,
\eqno(5.7)$$
we can easily prove the identity
$$    \widetilde{\phi}^{ML}_{\tau,a} (s)
= \widetilde{\phi}^{ML} (\tau s/a^{1/\beta })
\q \hbox{for all} \q \tau >0, \q a>0\,. \eqno(5.8)$$
Note that Eq. (5.8) states the {\it self-similarity} of  the combined operation
{\it rescaling-respeeding}
for the Mittag-Leffler waiting time density. In fact, (5.8) implies
$ {\phi}^{ML}_{\tau,a} (t) =
   {\phi}^{ML}(t/c) /c$ with
$ c= \tau /a^{1/\beta }\,,$
 which means replacing the random waiting time $T^{ML}$
by   $ c\, T^{ML}$.
As a consequences, choosing  $a= \tau ^\beta $ we have
$$    \widetilde{\phi}^{ML}_{\tau,\tau^\beta} (s)
 = \widetilde{\phi}^{ML} ( s)
\q \hbox{for all} \q \tau >0\,. \eqno(5.9)$$
Hence {\it the Mittag-Leffler waiting time density  is invariant against
    combined rescaling  with $\tau$ and respeeding with  
    $a=\tau^\beta$}.
	\vsp
	Observing  (5.6) we can say that $\phi^{ML} (t)$ is a $\tau \to 0$
	attractor for any power law waiting time
	(5.2) under simultaneous rescaling with $\tau$ and respeeding with
	$a= \lambda \tau^\beta$.
In other words, this attraction property of the Mittag-Leffler probability 
 distribution with respect to power law  waiting  times (with $0<\beta\le  1$) 
 is a kind of analogy to the attraction of sums of power law jump 
 distributions by stable distributions.

\section{Passage to the diffusion limit in space} 

We have again two different situations for the jump-width distribution
but according to its   second  moment being  finite or infinite.
In other words we assume for the jump-width probability density $w(x)$
(assumed for simplicity to be symmetric: $w(x)= w(-x)$)
either 
$$   \sigma ^2 := \int_{-\infty}^{+\infty} x^2\,w(x)\, dx
   < \infty\,, \q \hbox{labelled as} \; \alpha =2\,,\eqno(6.1) $$
   or
$$  w(x) \sim b\, |x|^{-(\alpha +1)}
 \q \hbox{for} \q |x|\to \infty\,, \q
0<\alpha <2 , \q b>0\,. \eqno(6.2) $$
Then
we have the asymptotic relation,
compare \eg with \cite{GAR_Vietnam04,GorMai_FCAA98,Gorenflo-Mainardi_INDIA03,Gorenflo-Mainardi_CARRY04},
$$  \widehat w(\kappa) = 1 -\mu\, {|\kappa|}^\alpha
    +o ({|\kappa|}^\alpha ) \q \hbox{for} \q \kappa  \to 0\,,
\eqno(6.3)$$
where 
$$ \mu = \frac{\sigma ^2}{2}\q \hbox{if} \quad \alpha=2,
\qq \mu= \frac{b\,\pi}{\Gamma(\alpha +1)\,
  \sin \left ( \alpha \pi/ 2 \right )}\q
 \hbox{if} \quad 0<\alpha<2,  \eqno(6.4) $$
 The above asymptotic relations are known in the framework of the
 attraction properties of the stable densities. We note   
 that the classical book by Gnedenko and Kolmogorov \cite{GnedenkoKolmogorov_54}
 has unfortunately the wrong constant $\mu$ for $0<\alpha<2$.
As before we dispense with the possible decoration
of the relevant power law by a slowly varying function.
\vsp
By another respeeding, in fact an acceleration, we can pass over to
space-time fractional
diffusion processes. For this we have
{\bf three choices}: 
\\
{\bf (a):  diffusion limit in space only, for general waiting time},
\\
{\bf (b):  diffusion limit in space only, for ML waiting time},
\\
{\bf (c): joint limit in time and space
(with power laws in both) with
scaling relation}.
\vsp
Note hat (b)  is just a special case of (a) but of particular relevance (as we shall see).
In all three cases we rescale the jump density by a factor
$h>0$, replacing the random     jumps $X$ by  $hX$.
This means changing the unit of measurement in space from $1$ to $1/h$,
 with $0 <h \ll 1$,
  so bringing into near sight the  far-away space.
We get the rescaled jump density as  $w_h(x) =  w(x/h)/h$,
corresponding to   $\widehat w_h(\kappa) =\widehat w_(h\kappa)$.
\vsp
{\bf Choice (a): diffusion limit  in space only, with  a general waiting time law}.
\\
Starting from the Eq. (3.6), the Laplace-Fourier representation of the CTRW equation,
 without special assumption on the waiting time density,
  we fix  the Fourier variable  $\kappa $
and accelerate the spatially rescaled
process by the respeeding factor $ 1/(\mu h^\alpha )$,
arriving at the equation (using  $q_h$   as new dependent variable)
$$   \widetilde{H} (s) \, \lt[
  s\widehat{\widetilde q}_h(\kappa ,s)-1\rt] =
 \frac{ \widehat w(h \kappa )-1}{\mu h^\alpha }\,
   \widehat{\widetilde q}_h(\kappa ,s)\,. \eqno(6.5) $$
Then, {\it fixing $\kappa$} as required
by the continuity theorem of probability theory for Fourier transforms,
and {\it sending $h$ to zero} we get, 
noting that $ [\widehat w(h \kappa )-1]/(\mu h^\alpha) \to -|\kappa|^\alpha$,
and
writing  $u$   in place of  $q_0$,
$$  \widetilde{H} (s) \, \lt[
  s\widehat{\widetilde u}(\kappa ,s)-1\rt] =
-   |\kappa |^\alpha
\,\widehat{\widetilde u}(\kappa ,s)\,, \eqno(6.6)$$
where we still have, consistently with (3.5), 
$$ \widetilde{H} (s) = \frac{1- \widetilde{\phi}(s) }
{ s\, \widetilde{\phi}(s)}
 = \frac{\widetilde{\Psi}(s) }{\widetilde{\phi}(s)}\,,$$ 
being $\phi (t)$ the original waiting time density.
In
physical space-time we have the integro-pseudo-differential equation
 $$ \int_0^t   H(t-t')\,
 \frac{\d}{\d t'} u(x,t')\, dt' \, =   \,
    _x D_{0}^{\,\alpha} \,u(x,t)
      \, ,\q 0<\alpha \le 2 \,,\eqno(6.7) $$
with $-|\kappa |^\alpha $
  as the symbol of the Riesz pseudo-differential operator
$\, _x D_{0}^{\,\alpha}$ usually referred to as the Riesz fractional
derivative of order $\alpha$, see Appendix B. 
\vsp
{\bf Comments}:
By this rescaling and acceleration the jumps become smaller and
smaller, their number in a given span of time larger and larger,
the waiting times between jumps smaller and smaller.
In the limit there are no waiting times anymore,
the original waiting
time density $\phi (t)$   is now only spiritual, but still determines via
$H(t)$ the memory of the process.
Eq. (6.7) offers a great variety of diffusion processes with memory depending
on the choice of the function $H(t)$.
\vsp
{\bf Choice (b): diffusion limit  in space only, with a Mittag-Leffler waiting time law}.
\\
We now choose in Eq. (6.7) the Mittag-Leffler memory function (3.17), namely  
$$  H(t) = H^{ML}(t)\,,  
\quad \hbox{hence} \quad \widetilde H(s)=1/s^{(1-\beta)}\,,$$
 corresponding to the Mittag-Leffler waiting time law 
$$    \phi^{ML} (t) =  - {\ds \frac{d}{dt} E_\beta (-t^\beta )}\,,
\q 0<\beta \le 1\,, $$
consistently 
with the time fractional Kolmogorov-Feller equation (3.15),
that includes for $\beta=1$ the classical   Kolmogorov-Feller equation (3.11).
As a consequence of our 
spatial diffusion limit, compare with \cite{Gorenflo_KONSTANZ01,Mainardi_PhysicaA00},   
we so arrive immediately at the 
{\it space-time  fractional diffusion equation}
$$  {\, _t}D_{*}^{\, \beta }\, u(x,t)
 \, = \,
 {\, _x}D_{0}^{\,\alpha} \,u(x,t)
\,,\quad 0<\alpha \le 2\,,\; 0<\beta \le 1\,.
\eqno(6.8)$$
\vsp
{\bf Choice (c): diffusion limit combined in time and space}.
\\
Assuming the  behaviour for the waiting time density as in  Eqs. (5.1)-(5.2), 
and for the jump-width density 
as in Eqs. (6.1)-(6.2),
rescaling as described the waiting times and the jumps by factors
$\tau $   and  $h$,
starting from
(4.4), decelerating by a factor $ \lambda\, \tau ^\beta $   in time,
 then accelerating for space by a factor $1/(\mu h^\alpha )$,
 we obtain (compare to Section  4, case (B)), fixing $s$ and $\kappa $
  and setting,
for convenience
$$ a(\tau ,h) = \frac{\lambda \tau ^\beta }{\mu h^\alpha } 
\, ,\eqno(6.9) $$
$$  \widetilde{H}_\tau  (s) \, \left[
  s\widehat{\widetilde {p}}_{\tau, a(\tau ,h)}(\kappa ,s)-1\right] =
 a(\tau ,h)\, \left[ \widehat w_h(\kappa) - 1\right]\,
   \widehat{\widetilde {p}}_{\tau, a(\tau ,h)} (\kappa ,s)\,,\eqno(6.10)$$
  with $\widehat w_h(\kappa) = \widehat w(h\kappa)$ and 
$$
\widetilde{H}_\tau (s)=
   \frac{1- \widetilde{\phi}(\tau s) }
{ s\, \widetilde{\phi}(\tau s)} \sim \lambda \tau ^\beta s^{\beta -1 }\,.\eqno(6.11)$$
Fixing $a(\tau,h)$ to the constant value 1, which means  introducing the relationship
of {\it well-scaledness}
$$ a(\tau,h)= \frac{\lambda \tau ^\beta}{\mu h^\alpha } \equiv 1\,, \eqno(6.12)$$
between the rescaling of time and space, we get
$$ \widetilde{H}_\tau (s) \sim \lambda \tau^\beta\, s^{\beta-1}\,,
\quad \hbox{for} \quad \tau \to 0\,.\eqno (6.13)$$
Because of  
$$\frac{ \widehat w(h \kappa )-1}{\mu h^\alpha } \to -|\kappa|^\alpha\,,
\quad \hbox{for} \quad h \to  0\,,\eqno(6.14)$$
we finally get
   the limiting equation
  $$
  s^{\beta -1} \, \lt[
  s\widehat{\widetilde u}(\kappa ,s)-1\rt] =
- |\kappa |^\alpha  \,
   \widehat{\widetilde u}(\kappa ,s)\,,\eqno(6.15)  $$
 corresponding to Eq. (6.8), the space-time  fractional
diffusion equation.
\vsp
{\bf Comments on some mathematical and physical aspects}
\vsp 
{\bf ($\alpha$)} The Mittag-Leffler waiting time
(choice (b)), obeying the power law
asymptotics (5.2) with $\lambda =1$   leads from (6.7) directly to the space-time  fractional
diffusion equation (6.8), without requirement of rescaling and deceleration in time, and
with these procedures
we arrive likewise at (6.8). This strange fact is caused by the invariance
of the Mittag-Leffler density to the combined effects of
rescaling by  $\tau $ and deceleration by  $\tau ^\beta$,
expressed in eq. (5.9).
\vsp 
{\bf ($\beta$)}
 Going again through our preceding deductions, we observe that
 the combined (well-scaled) passage of $\tau$ and $h$, 
 under the relation (6.12),
 towards zero can be split
 in two distinct ways into two separate passages.
 \underbar{First way}: keep $h$ fixed letting $\tau$ tend to zero, then
 in the resulting model send also $h$ to zero.
 \underbar{Second way}: interchange the order played 
 by $h$ and $\tau$ in the first way. Under our power 
 law assumptions we can transform (3.7), the basic integral equation of 
 CTRW,  into Eq. (3.15) (time fractional CTRW) by rescaling-respeeding 
 manipulation only in the time variable, and then by rescaling in space 
 followed by an acceleration into (6.8), the space-time fractional 
 diffusion equation. Or we can transform (3.7) by rescaling in space 
 followed by an acceleration into Eq. (6.7) (general space fractional 
 diffusion with memory), and then by by rescaling-respeeding in the 
 time variable arrive at (6.8).
 \vsp 
 {\bf ($\gamma$)} Where have the waiting times gone in the space-time
 fractional diffusion equation (6.8)?
 We can answer this question by interpreting eq. (6.10) under the scaling relation (6.12) as
 the Laplace-Fourier representation 
$$  \widetilde{H}_\tau  (s) \, \left[
  s\widehat{\widetilde {p}}_{\tau, 1}(\kappa ,s)-1\right] =
 \left[ \widehat w_h(\kappa) - 1\right]\,
   \widehat{\widetilde {p}}_{\tau, 1} (\kappa ,s)\,,\eqno(6.16)$$
   of our original CTRW (3.1), whose
   Laplace-Fourier representation (3.6) coincides with (6.16) 
   if there we delete all decorations with indices.
   Thus eq. (6.16) represents the same physical process as (3.1) but expressed in terms of new units
   $1/\tau$ and $1/h$ of time and space, respectively. 
   However, the respeeding factor $a(\tau,h)$ being fixed to 1,
   there is no change of physical speed.
   When these new units  are made smaller and smaller, moderate spans of time and space
   become numerically smaller and smaller, shrinking towards zero
   as $\tau$and $h$ tend to zero, and likewise the waiting times and the jump widths
   shrink to zero.
   The distant future and the far-away space come numerically into near sight.
   As long as $\tau$ and $h$ are positive, we always have the same physical process,
   only measured in other units.
   The finally resulting  space-time fractional diffusion process (6.8)
   remembers the power laws for waiting times and jumps in form
   of the orders $\beta$ and $\alpha$ of fractional differentiation.
 \vsp 
 {\bf ($\delta$)}
An objection could be raised against the somewhat mystical actions of respeeding.
Namely, if the respeeding factor $a$ in eq. (4.4) differs from 1, the underlying renewal process and
consequently the whole CTRW are distorted. However, for the CTRW we carry out the actions of
deceleration and acceleration in either order in succession or simultaneously in combination,
and by our special choice of these factors they
cancel each other in effect, so that there remains no physical distortion.
This is particularly obvious in our choice (c), see the above comment ($\gamma$). 
\vsp 
{\bf ($\epsilon$)}
Let us finally point out an advantage of splitting the passages
$\tau \to 0$ and $h \to 0$.
Whereas by the combined passage as in choice (c), if done in the well-scaled way (6.12),
the mystical concept of respeeding can be avoided, there arises
the question of correct use of the continuity theorems
of probability. There is one continuity theorem for the Laplace transform,
one for the Fourier transform, see \cite{Feller_71}.
Possible doubts whether their simultaneous use is legitimate
vanish by applying them in succession, as in our
two splitting methods.
\vsp
{\bf Discussion on the involved stochastic processes}
\vsp
In our   investigations we have met  four types of
spatially one-dimensional stochastic processes 
for the sojourn probability density $p(x,t)$ or $u(x,t)$.
For the reader's convenience let us give a list of these processes
in physical coordinates, referring to the preceding text for details, 
and remind briefly how they
can be connected by appropriate scaling and passages to the limit.
Let us note that in all these processes the initial condition
 $\delta(x)$ for  $p(x,0^+)$ or $u(x,0^+)$  can be replaced by a more general 
 probability density function $f(x)$.
\vsp {\bf (I)} 
The integral equation for the CTRW is, see  (3.1) with (3.2),
$$
   p(x,t) =  p(x,0^+)\, \Psi(t)\, +
  \int_0^t  \!\!  \phi(t-t') \, \lt[
 \int_{-\infty}^{+\infty}\!\!  w(x-x')\, p(x',t')\, dx'\rt]\,dt' $$
is equivalent,  by the introduction of the  memory function $H(t)$, see (3.5),
to the generalized Kolmogorov-Feller equation, see (3.7),
$$ \int_0^t   H(t-t')\,
 \frac{\d}{\d t'} p(x,t')\, dt'  =
      -p(x,t) + \int_{-\infty}^{+\infty} w (x-x')\,p(x',t)\,dx'\,.
$$
\vsp {\bf (II)} 
 The time fractional Kolmogorov-Feller equation, see (3.15),
$$   \, _tD_*^\beta \,  p(x,t) =
     -  p(x,t) +   \int_{-\infty}^{+\infty} w(x-x')\,
   p(x',t) \, dx'\,, \quad 0<\beta \leq 1\,.  $$
\vsp {\bf (III)}
The integro-pseudo-differential equation  of space fractional diffusion
with general memory, see (6.7), 
$$ \int_0^t   H(t-t')\,
 \frac{\d}{\d t'} u(x,t')\, dt' \, =   \,
    _x D_{0}^{\,\alpha} \,u(x,t)
      \, ,\q 0<\alpha \le 2 \,. $$
\vsp 	{\bf(IV)}	 
The space-time fractional diffusion equation, see (6.8),
  $$  {\, _t}D_{*}^{\, \beta }\, u(x,t)
 \, = \,
 {\, _x}D_{0}^{\,\alpha} \,u(x,t)
 \quad 0<\beta \leq 1\,, \quad 0<\alpha \le 2
\,.$$
\vsp   
We 	now sketch shortly how these four evolution equations are
connected in our theory.
Eq. (I) goes over in eq. (II), likewise eq. (III) in eq. (IV)
by the special choice $H(t)= H^{ML}(t)$ for the memory function, see (3.17).
Under our power law assumption for the waiting time, see (5.1) and (5.2),
these transitions can be achieved asymptotically by manipulation
via rescaling and respeeding of the underlying renewal process.
Under our power law assumption for the jumps,
see (6.1)and (6.2), the transition from
eq. (I) to eq. (III)  and from (II) to (IV)
can be achieved asymptotically by passage  to the diffusion limit
only in space.
Under our power law assumption for time \underbar{and} space there is
a direct way  from eq. (I) to eq. (IV), namely {\it the well-scaled
passage to the diffusion limit}, for which the condition (6.12) is relevant.

\section{The time fractional drift process}  

It is instructive to study the spatial transition to the diffusion limit
for the {\it Mittag-Leffler renewal process}.
As said in Section 3 this renewal process, viewed as a CTRW
by treating its counting number $N$ as a spatial variable $x$, is obtained by choosing
$ w(x) =\delta (x-1)$
    as the jump width density, see Eq. (3.3). 
	Its waiting time density is, see (1.1), (3.13),
$$  \phi (t) = \phi^{ML} (t) =  - {\ds \frac{d}{dt} E_\beta (-t^\beta )}\,,
 \q 0<\beta \le 1\,.$$
   We have
$\widetilde H(s) = s^{\beta -1}$, $\,\widehat w(\kappa ) = \e^{i\kappa}$,
hence
$$ s^{\beta -1}\lt [ s\widehat{\widetilde {p}}(\kappa ,s)-1\rt] =
  \lt( \e^{i\kappa}-1\rt)\,
   \widehat{\widetilde {p}} (\kappa ,s)
\,. \eqno(7.1) $$
Rescaling in space by a factor $h$  and accelerating (because of
$  w(\kappa ) = \e^{i\kappa}= 1 +i\kappa  + o(\kappa )$
for $\kappa \to 0$)
  this pure renewal process by the factor $1/h$    we get a process
$$ s^{\beta -1}\lt [ s\widehat{\widetilde {q}_h}(\kappa ,s)-1\rt] =
 \rec{h} \lt( \e^{i h \kappa}-1\rt)\,
   \widehat{\widetilde {q}_h} (\kappa ,s)
\,,  $$
   which as  $h\to 0$ and $\kappa $ fixed   gives
$$  s^{\beta -1}\lt [ s\, \widehat{\widetilde {u}}(\kappa ,s)-1\rt] =
 i\kappa \,
   \widehat{\widetilde {u}} (\kappa ,s)
\,,  \eqno(7.2) $$
which implies
$$
\widehat{\widetilde {u}}(\kappa ,s)  =
\frac{ s^{\beta -1}}{s^\beta -i\kappa}\,.
\eqno(7.3)$$
We note that Eq. (7.2)
corresponds to the  time fractional drift equation
$$   {\, _t}D_{*}^{\, \beta }\, u(x,t)
 \, = \,
  - \frac{\d}{\d x} u(x,t)
\,,\quad  u(x,0) = \delta (x)\,,
\q x\in \RR^+\,,\; t \in \RR^+\,.
 \eqno(7.4)$$
By using the known scaling rules
for the Fourier  and Laplace  transforms,
$$   f(ax) \,\stackrel{{\F}} {\leftrightarrow} \,
  a^{-1}\, \widehat   f(\kappa/a)\,,\q   a> 0\,,
 \qq   f(bt) \,\stackrel{{\L}} {\leftrightarrow}\,
 b^{-1}\, \widetilde f(s/b)\,, \q b> 0\,,$$
we  infer directly from (7. 3) (thus without inverting the
two transforms)   the  following {\it scaling property}
of the (fundamental) solution
$$    u(ax\,,\,bt)
= b^{-\beta } u (ax/b^\beta\,,\,t)\,.
    $$
Consequently, introducing the {\it similarity variable} $x/t^\beta \,,$
we  can write
$$ u(x,t)  =
    t^{-\beta }\,U (x/t^\beta )
\,,   \eqno(7.5)     $$
where $U(x) \equiv u(x,1)\,. $
\vsp
To determine the solution
in the space-time domain we can follow two
alternative  strategies related to the  different
order in carrying out the
inversion of the Fourier-Laplace transforms
in (7. 3).
Indeed we can
\\
(S1) : invert  the Fourier transform
getting $\widetilde{u} (x,s)\,,$
   and    then invert this Laplace transform,
\\
(S2) : invert  the Laplace transform
getting
$\widehat{u} (\kappa ,t)\,,$
and then invert this Fourier transform.
\vsp
{\it Strategy (S1):}
Recalling
the Fourier transform pair,
$$  
  a  \,\e^{\ds \,- x b}\,\Theta(x)
 \,\stackrel{{\F}} {\leftrightarrow}\,
 \frac{a}{ b - i\kappa }\,,
 \q b >0 \,,
$$
where $ \Theta(x)$ denotes the unit step Heaviside function,
we get
$$ \widetilde{u} (x,s) =
 s^{\beta-1}\,\e^{\ds \,- x s^{\beta}}\,\Theta(x)
\,.\eqno(7.6) $$
In view  of the fact that $\exp{(-s^\beta)}$
is the Laplace transform of the extremal unilateral stable density of order $\beta$,
$L_\beta^{-\beta}(t)$ (see for notation Appendix B), we  recognize that the solution
in the space-time domain can be expressed in terms of a fractional integral
(see Appendix A)  of such density, namely
 $$   u(x,t) = \frac{1}{ x^{1/\beta}}\; _tJ^{1-\beta}\, \left[L_\beta^{-\beta} \left( {t}/{x^{1/\beta}} \right)\right]\,.
\eqno(7.7)
$$
Working in the Laplace domain
we can note that the fundamental solution of our fractional drift equation (7.4)
is simply related to that of 
 the time fractional diffusion-wave  equation 
$$ \,_tD_*^{2 \beta} u(x,t) = \frac{\d^2}{\d x^2} u(x,t)\,, \q 0<\beta \le 1\,, 
\q  x\in \RR\,,\; t \in \RR^+\,,
\eqno(7.8)$$
equipped with the initial conditions
 $u(x,0^+) = \delta (x)$ if $0<\beta\le 1$ and ${\ds \frac{\d}{\d t} u(x,0^+)= 0}$ if
  $1/2<\beta \le 1$.
In fact, the solution of (7.8)
turns out  the half of the solution (7.7) of our time fractional drift equation (7.4),
extended in a symmetric way to all of $\RR$, as can be
seen by factorizing eq. (7.8) as
$$  \left(\,_tD_*^{2 \beta} -\frac{\d^2}{\d x^2}\right)\, u(x,t)\, =\,
\left(\,_tD_*^{\beta} - \frac{\d}{\d x}\right) \,\left(\,_tD_*^{\beta} + \frac{\d}{\d x}\right) \,
u(x,t) =0\,.
\eqno(7.8')$$
Indeed eq. (7.8) was solved 
by using the Laplace transform strategy 
by Mainardi  in the 1990's,
see e.g. \cite{Mainardi_CHAOS96,Mainardi_CISM97,Mainardi-Pagnini_AMC03}
where the reader can find mathematical details of the proof and 
instructive plots of the fundamental solution.   
Then, based on Mainardi's analysis, we can state
that the required solution of eq. (7.4) reads
  $$ u(x,t) = t^{-\beta } \, M_\beta (x/t^\beta)\, \Theta (x)\,,
\eqno(7.9)
$$
where $M$ denotes
the function of Wright type
defined in the complex plane
$$ M_\beta  (z) \!=\!
  \sum_{n=0}^{\infty}\,
  {(-z)^n\over n!\,\Gamma[-\beta  n + (1-\beta )]} 
  \!=\!
  \rec{\pi}\, \sum_{n=1}^{\infty}\,{(-z)^{n-1} \over (n-1)!}\,
  \Gamma(\beta  n)  \,\sin (\pi \beta   n).    \eqno(7.10)     $$
The $M$ function  is a special case of the
Wright function  defined by the  series  representation,
 valid in the whole complex plane,
  $$ \Phi _{\lambda ,\mu }(z ) :=
   \sum_{n=0}^{\infty}{z^n\over n!\, \Gamma(\lambda  n + \mu )}\,,
 \q \lambda  >-1\,, \,\q \mu \in \CC\,, \q z \in \CC \,.
\eqno(7.11)
$$
Indeed, 
 we  recognize
   $$ M_\beta  (z) =  \Phi _{-\beta  , 1-\beta  }(-z) \,, \q 0<\beta  <1\,.
  \eqno(7.12)  $$
Originally, Wright introduced and investigated this function
with the restriction $\lambda \ge 0$ in a series of notes starting from
1933 in the framework of the asymptotic theory of partitions. Only later,
in 1940, he considered the case $-1<\lambda <0$.
We note that in the handbook of the Bateman Project
\cite{Erdelyi_HTF} (see Vol. 3,  Ch. 18),
presumably for a misprint,  $\lambda $ is
restricted to be non negative.
 For further mathematical details on the $M$-Wright function 
 we recommend \cite{GoLuMa_99,GoLuMa_00,GorMaiSri_PLOVDIV97}.
\vsp
For our time fractional drift equation (7.4) we note the  particular case
  $\beta =1/2$
for which we obtain
$$ \beta =1/2\,: \q
u(x,t) =\rec{\sqrt{\pi t}} \,\exp \lt[-x^2/(4t)\rt]\,, \q x\ge 0 \,,\q t\ge 0\,.
\eqno(7.13)$$
In the limiting case $\beta=1$ we recover the rightward pure drift,
$$ \beta =1\,: \q
u(x,t) =\rec{t} \, \delta (x/t -1) = \delta (x-t)\,,
\q x\ge 0 \,,\q t\ge 0\,.\eqno(7.14)$$
In view  of the fact that 
that $M$-Wright function of order $\beta$
is related to the extremal unilateral stable density of order $\beta$,
see  \cite{Mainardi_LUMAPA01},
we conclude by displaying the alternative form of the solution of the 
time fractional drift equation: 
$$   u(x,t) = \frac{t}{\beta \, x^{1+1/\beta}}\, L_\beta^{-\beta} \left( {t}/{x^{1/\beta}} \right)\,,
\eqno(7.15)
$$
which, compared with (7.7), shows the effect of the fractional
integral on the stable density function $L_\beta^{-\beta}$. 
\vsp
{\it Strategy (S2):}
Recalling
the Laplace transform pair,
see e.g.    \cite{Erdelyi_HTF,GorMai_CISM97},    
$$
 E_{\beta} (ct^\beta)
  \,\stackrel{{\L}} {\leftrightarrow}\,
\frac{ s^{\beta-1}}{s^\beta -c}  ,
 \q \Re \,(s) > |c|^{1/\beta}\,,
 $$
we get
$$ \widehat{{u}}(\kappa ,t) = E_\beta (i \kappa  t^\beta )\,, \eqno(7.16)$$
from which
$$ u(x,t) = \frac{1}{2\pi} VP \int_{-\infty}^{+\infty}
  \e^{\ds\,-i\kappa x}\, E_\beta(i \kappa  t^\beta )\,d\kappa \,,
\eqno(7. 17) $$
where $VP$ denotes the Cauchy principal value.
Because, see \cite{Erdelyi_HTF},
 Vol. 3, Chapter XVIII on Miscellaneous Functions,
Section 18.1 Eq. (7), 
$$E_\beta (iy) \sim \frac{i}{\Gamma(1-\beta)y}
\q \hbox{for} \q  y \to \pm \infty\, 
\q \hbox{if} \q 0<\beta<1\,,$$
 we see
that $E_\beta (iy)$ does not tend to zero fast enough for
the integral (7.17) to exist as a regular improper Riemann integral.
But there should be no problem for existence as a Cauchy principal
value integral.
It can be shown that the present strategy based on Fourier integral (7.17)
provides the  result (7.9).  
\vsp
{\bf Remark}:
Not wanting to overload our paper we have deliberately avoided
the concept of {\it subordination} in fractional diffusion.
But, referring to \cite{GorMaiViv_CSF06},
let us say that if in (7.15) we replace $x$ by $t_*$ we get  the
{\it subordinator}, i.e. the probability law for generating the operational time $t_*$
from the physical time $t$, see eq. (5.20) in  \cite{GorMaiViv_CSF06},
and, in other notation, \cite{M3_PRE02sub}.
Because of its relation (7.16) via Fourier transform to the
Mittag-Leffler function with imaginary argument,
 the probability law governing the process (7.15) sometimes is called
the {\it Mittag-Leffler distribution}, see e.g. \cite{M3_PRE02sub}.
Although so named it must not be confused
with our {\it Mittag-Leffler waiting time distribution}
whose density is given by (1.1).

\section{Conclusions}
 The basic role of the Mittag-Leffler waiting time probability density in time fractional continuous time 
 random walk (CTRW) has become well known by the fundamental paper of 1995 by 
 Hilfer and Anton \cite{Hilfer-Anton_PRE95}.
 Earlier in the theory of thinning (rarefaction) of a renewal process under power law assumptions, 
 see the 1968 book by Gnedenko and Kovalenko \cite{GnedenkoKovalenko_QUEUEING68},  
 this density had been found as limit density by a combination of thinning followed 
 by rescaling of time and imposing a proper relation between the rescaling factor and the thinning parameter. 
 Likewise one arrives at this law when wanting to construct a certain special class of anomalous random walks,
 see the 1985 paper  by Balakrishnan \cite{Balakrishnan_85}, 
 the anomaly defined by growth of the second moment of the sojourn probability density 
 like a power of time with exponent between 0 and 1.
Balakrishnan's paper, having appeared a few years before the fundamental paper of 1989 
by Schneider and Wyss \cite{SchneiderWyss_89}, 
is difficult to read as it is written in a style different from the present one, so we will here not go into details. 
But let it be said that by well-scaled passage to the limit from CTRW (again under suitable power law assumptions 
in space and time) he obtained the space-time fractional diffusion equation in form of 
an equivalent integro-differential equation.
Unfortunately, Balakrishnan's paper did not find the attention it would have deserved.
However, due to the sad fact that the Mittag-Leffler function too long played a rather neglected role in treatises 
on special functions Balakrishnan as well as Gnedenko and Kovalenko contented themselves with presenting their 
results only in the Laplace transform domain; 
they did not identify their limit density as a Mittag-Leffler type function.  
\vsp
Having worked ourselves for some time on questions of well-scaled passage to the diffusion limit from continuous
 time random walks to fractional diffusion, 
 see \cite{Gorenflo-Mainardi_INDIA03,Gorenflo-Mainardi_CARRY04,Gorenflo_KONSTANZ01,GorVivMai_NLD04,Mainardi_FNL05,Scalas_PRE04},
 we got from the theory of thinning the idea that it should be possible 
 to carry out the passages to the limit separately in space and in time. 
 In time this can be done by a combination of re-scaling time and respeeding the underlying renewal process 
 (formally treating it as a CTRW with unit steps in space).
In fact, {\it thinning} in the sense of Gnedenko and Kovalenko transforms the original renewal process into one that 
is running more slowly and this effect can be balanced by proper choice of the rescaling factor. 
The result of our combination of rescaling and respeeding for a  CTRW governed by a given renewal process with a 
generic power law waiting time law is a time fractional CTRW. 
By another rescaling in space (now under power law assumption for the
jumps) which can be interpreted as a second respeeding we arrive at the already classical space-time fractional 
diffusion equation. In this way we shed new light on the long time and wide space behaviour of 
continuous time random walks.
\vsp
In a series of comments at the end of Section 6, we have explained  how, by what we call
{\it well-scaled} passage to the diffusion limit, the transition from the CTRW to the
space-time fractional diffusion process actually can be obtained by merely rescaling time and space
without any respeeding at all. However, the separate passages to the limit are more
satisfying with respect to mathematical rigour.
\vsp
Finally, in Section 7, we have treated the time fractional drift process as a properly scaled limit of the 
counting function   of a pure renewal process governed by
a waiting time law of Mittag-Leffler type.
 Our trick in finding the limiting waiting time law of this renewal process
consists in  treating it
 as a CTRW with positive jumps of size $1$ so that its counting number acts as a spatial variable. 
Then, by suitably rescaling this spatial variable,
we obtain as an interesting side 
result  the long time behaviour of the  Mittag-Leffler renewal process.   
\section*{Appendix A: The time fractional derivatives}
  
   For a sufficiently well-behaved function $f(t)$ ($t\ge 0$) we
 define the {\it Caputo time fractional derivative}  of
order $\beta  $  with $0<\beta <1$
through
$$ {\cal L} \left\{ _tD_*^\beta \,f(t) ;s\right\} =
      s^\beta \,  \widetilde f(s)
   -s^{\beta  -1}\, f(0^+) 
   \,, \quad f(0^+):= \lim_{t \to 0^+} f(t)\,,
    \eqno(A.1)$$ 
 so that
 $$   _tD_*^\beta \,f(t) :=
\rec{\Gamma(1-\beta )}\,\int_0^t
 {f'(\tau)\over (t-\tau )^\beta} \, d\tau\,, \q 0<\beta<1 \,.
 \eqno(A.2)
  $$
Such operator  has been referred to as
the {\it Caputo} fractional derivative since it
was introduced by Caputo in the late 1960's
for modelling the energy dissipation
 in the rheology of the Earth, see  \cite{Caputo_67,Caputo_69}.
  Soon later this derivative was adopted by Caputo and Mainardi
  in the framework of the linear theory of viscoelasticity, see \cite{CaputoMaina_71}.
\vsp
The reader should observe that the {\it Caputo} fractional derivative
differs from the usual {\it Riemann-Liouville} (R-L) fractional derivative
$$
 \,_tD^\beta  \,f(t) :=
  {\ds {d\over dt}}\,\left[
  {\ds \rec{\Gamma(1-\beta )}\,\int_0^t
    {f(\tau)\,d\tau  \over (t-\tau )^{\beta  }} }\right] \,, \q 0<\beta<1 \,.
	\eqno(A.3)
  $$
 Both derivatives are related to the  Riemann Liouville (R-L) fractional integral
 that is defined for any order $\beta>0$ as
$$\, _tJ^\beta  \,f(t) :=
  {\ds \rec{\Gamma(\beta )}\,\int_0^t
    \frac{f(\tau)\,d\tau } {(t-\tau )^{1-\beta  }} }\,, \q \beta >0\,,\eqno(A.4)
  $$    
 so that  $ {\cal L} \left\{ _tJ^\beta \,f(t) ;s\right\} =
      s^{-\beta} \,  \widetilde f(s)$.
Incidentally  $\,_tJ^\alpha \,_tJ^\beta \,= \,_tJ^{\alpha+\beta}  $ for  $\alpha, \beta >0$.	  
Then, in virtue of eqs (A.2)-(A.4), the two fractional derivatives read: 
$$ \,_tD^\beta  :=  \,_tD^1\,_tJ^{1-\beta}\,, \q 0<\beta<1 \,, \eqno(A.5)$$ 
  $$\,_tD_*^\beta  :=  \,_tJ^{1-\beta}\,_tD^1\,,\q 0<\beta<1 \,. \eqno(A.6)$$ 	  
  In particular, the R-L derivative of order $\beta$
	 is  the left inverse of the corresponding R-L fractional integral
in that $\,_tD^\beta  \, _tJ^\beta \, f(t) = f(t)$.
\vsp
  We note the  relationships between the two fractional
derivatives  (when both of them exist),
for $0<\beta<1$,
$$ _tD_*^\beta	\,f(t)	
\, = \, _tD^\beta  \,\left[ f(t) -
   f(0^+) \right]
   	\, = \, _tD^\beta  \, f(t) -
     \frac{t^{-\beta }} {\Gamma(1-\beta)}\,f(0^+)  \,. 
  \eqno(A.7) $$
As a consequence we can interpret the Caputo derivative as   a sort of regularization of the R-L derivative
as soon as $f(0^+)$ is finite; in this sense such fractional derivative was independently introduced
in 1968 by     Dzherbashyan and Nersesian  \cite{Dzherbashyan-Nersesian_68},
as pointed out by Kochubei, see \cite{Kochubei_89,Kochubei_90}.
In this respect the regularized fractional derivative is sometimes referred to as the
{\it Caputo-Dzherbashyan derivative}.
\vsp
We observe the different behaviour  of the two fractional derivatives (A.2), (A.3)
at the end points of the parameter interval $(0,1)$,
 as it can be noted from their 
definitions in operational terms (A.5), (A.6). 
In fact, whereas for $\beta \to 1^-$ both derivatives 
reduce to $\,_tD^1$,  
 due  to the fact that the operator $\,_tJ^0 =\, I$ commutes with $\,_tD^1$,
 for $\beta \to 0^+$ we have
$$ 
 \beta \to 0^+     
 \Longrightarrow
 \cases{
 {\ds \,_tD^{\beta} f(t)} \to   {\ds \,_tD^1 \, _tJ^1 \, f(t) =  f(t)}\,,\cr
  {\ds \,_tD_*^{\beta} f(t)} \to  {\ds \,_tJ^1\, \,_tD^1 \, f(t)} =
{\ds f(t) - f(0^+) } \,.}
  \eqno(A.8)
$$
The above  behaviours have induced us to keep for the Riemann-Liouville
derivative the same symbolic notation as for the standard derivative of integer order,
while for the Caputo derivative to decorate the corresponding symbol with  subscript $*$.  
\vsp
For the R-L derivative the Laplace transform reads for $0<\beta<1$
$$ {\cal L} \left\{ _tD^\beta \,f(t) ;s\right\} =
      s^\beta\,  \widetilde f(s) - g(0^+)\,, \;
   g(0^+)= \lim_{t \to 0^+}\,g(t)\,, \; g(t):=\,_tJ^{(1-\beta)}\,f(t)  \,.\eqno(A.9) $$
    Thus the rule  (A.9) is     more cumbersome to be used
than (A.1) since it requires the initial value  of
an extra function $g(t)$ related to the given $f(t)$  
through a  fractional integral.
However, when $f^(0^+)$  is finite   we recognize $g(0^+)=0$.
\vsp   
  In the limit $\beta \to 1^-$ both derivatives reduce to the  derivative of
  the first order so we recover the corresponding standard formula for the 
  Laplace transform:
    $$ {\cal L} \left\{ _tD^1 \,f(t) ;s\right\} =
      s\,  \widetilde f(s) - f(0^+)\,. \eqno(A.10) $$
  We conclude this Appendix noting that
  in a proper way both derivatives can be generalized for any order $\beta>1$, see 
  \eg \cite{GorMai_CISM97,Podlubny_BOOK99}.

\section*{Appendix B: The space fractional derivatives}

Let us first recall that
a generic linear pseudo-differential operator $A$,
acting with respect to the variable $x \in \RR\,,$
is defined through its Fourier representation, namely
  $$   
  {\mathcal F} \left\{A \, f(x);\kappa \right\} :=
   \int_{-\infty}^{+\infty}
  \e ^{\, i\kappa x} \,  A \, f(x) \, dx =
 \widehat A(\kappa )\, \widehat f (\kappa )\,,\quad \kappa \in \RR \eqno(B.1)  $$
  where
$\widehat A(\kappa)\,$ is referred to as the symbol of $A$,
formally    given as
 $$ \widehat A (\kappa ) = \left( A\, \e^{\, -i\kappa x}\right)\,
  \e^{\, +i\kappa x}\,. \eqno(B.2) $$
The fractional {\it Riesz}  derivative
$\,_xD_0^\alpha   $  is defined
as the pseudo-differential operator  with symbol $-|\kappa |^\alpha\,. $
This means that for a sufficiently well-behaved (generalized)
function $f(x)$ ($x\in \RR$)
we have
$$ {\mathcal F} \left\{\,\,_xD_0^\alpha \, f(x);\kappa \right\} =
  - |\kappa| ^\alpha   \,
 \widehat f(\kappa) \,,  \q \kappa \in \RR\,. \eqno(B.3)
$$
The symbol of the Riesz fractional derivative is nothing but the
logarithm of the characteristic function of the generic symmetric
 {\it stable}
(in the L\'evy sense)
probability density, see
\cite{Feller_52,Feller_71, SKM_93}.   Noting
$    -|\kappa |^\alpha  = - (\kappa ^2)^{\alpha /2}\,,$
we recognize that
$$ {\ds \,_xD_0^\alpha  = - \left(-{d^2\over dx^2}\right) ^{\alpha/2}}\,. \eqno(B.4)$$
In other words, the Riesz derivative is a symmetric fractional
generalization of the second derivative to orders less than 2.
In an explicit way  the Riesz derivative reads, for $0<\alpha<2$,
$$  
\begin{array}{ll}
_xD_0^\alpha &\!\!f(x)  
= \,{\ds \frac{d^\alpha}{d|x|^\alpha} \,f(x)} \\
&= \,{\ds \Gamma(1+\alpha )
 \frac{\sin (\alpha \pi/2)}{ \pi } 
 \int_0^\infty 
 \frac{f(x+\xi)- 2f(x) + f(x-\xi)}{\xi^{1+\alpha}}\, d \xi}\,,  
\end{array}
 \eqno(B.5)$$
 where in the L.H.S
 we have also adopted  the alternative and  illuminating notation introduced by Zaslavsky,
see \eg \cite{Saichev_CHAOS97}.
This operator is  referred to as
the {\it Riesz fractional derivative} since it is obtained from the
inversion of the fractional integral originally introduced by
Marcel Riesz in the late 1940's, known as the {\it Riesz potential},
see \eg \cite{SKM_93}.  
It is based on a suitable regularization of a
hyper-singular integral, according to a method formerly
introduced by Marchaud in 1927.
\vsp
{\bf Remark}:
Straightforward generalization to the {\it Riesz-Feller derivative}
 of order $\alpha$ and skewness $\theta$ is possible.
Such pseudo-differential operator is denoted by us as  
$$  _x D_{ \theta}^{\,\alpha}\,, \q \hbox{with }\q 0<\alpha \le 2\,,
\q \theta \in \RR,\q |\theta| \le \hbox{min}\{\alpha,2-\alpha\}\,.\eqno(B.6)$$
In this case we have
$$ {\mathcal F} \left\{\,\,_xD_\theta^\alpha f(x);\kappa \right\} =
   - |\kappa |^{\,\ds \alpha}  \, i ^{\,\ds \theta \,\sgn \kappa} \,
   \widehat {f}(\kappa )
  \q \kappa \in \RR 
   \,. \eqno(B.7)$$
In an explicit way  the Riesz-Feller  derivative reads, for $0<\alpha<2$,    
$$
\begin{array}{ll}
_xD_\theta^\alpha \,f(x)
 &= \,{\ds {\Gamma(1+\alpha) \over \pi } \,
 \left\{\sin \,[(\alpha+\theta) \pi/2] \,
 \int_0^\infty
 {f(x+\xi)- f(x)  \over {\xi}^{1+\alpha}}\, d \xi \right.}\\
&+ \,{\ds \left. \sin \,[(\alpha-\theta) \pi/2] \,
 \int_0^\infty
 {f(x-\xi)- f(x)  \over {\xi}^{1+\alpha}}\, d \xi \right\}}\,.
 \end{array}
 \eqno (B.8)$$
Notice in (B.7) that
$i ^{\,\ds \theta \,\sgn \kappa}= \exp [ i \,(\sgn \kappa)\,\theta\,\pi/2 ]$.
Thus the symbol of the Riesz-Feller fractional derivative is the
logarithm of the characteristic function of the more general
(strictly) {\it stable}
probability density, closely following the Feller parameterization,
see \cite{Feller_52,Feller_71} revisited by 
the present authors  in \cite{GorMai_FCAA98}.
According to our notation, the strictly stable density of order $\alpha$
and skewness $\theta$ is denoted by $L_\alpha^\theta(x)$.
  We note that the allowed region for the 
parameters $\alpha $ and $\theta$
turns out to be
 a {diamond} in the plane $\{\alpha,\, \theta\}$
with vertices in the points
$(0,0)$, $(1,1)$, $(2,0)$, $(1,-1)$,
that we call the {\it Feller-Takayasu diamond},
see Fig. 1.
For more details  we refer the reader to \cite{Mainardi_LUMAPA01},
where series representations  and numerical plots 
of the stable densities $L_\alpha^\theta(x)$ are found.
In particular, we recall that the extremal stable densities obtained for
$\theta= \pm \alpha$ with $0<\alpha<1$ are unilateral,
with support in $\RR^{\mp}$, respectively.
\newpage
\begin{center}
\includegraphics[width=0.40\textwidth]{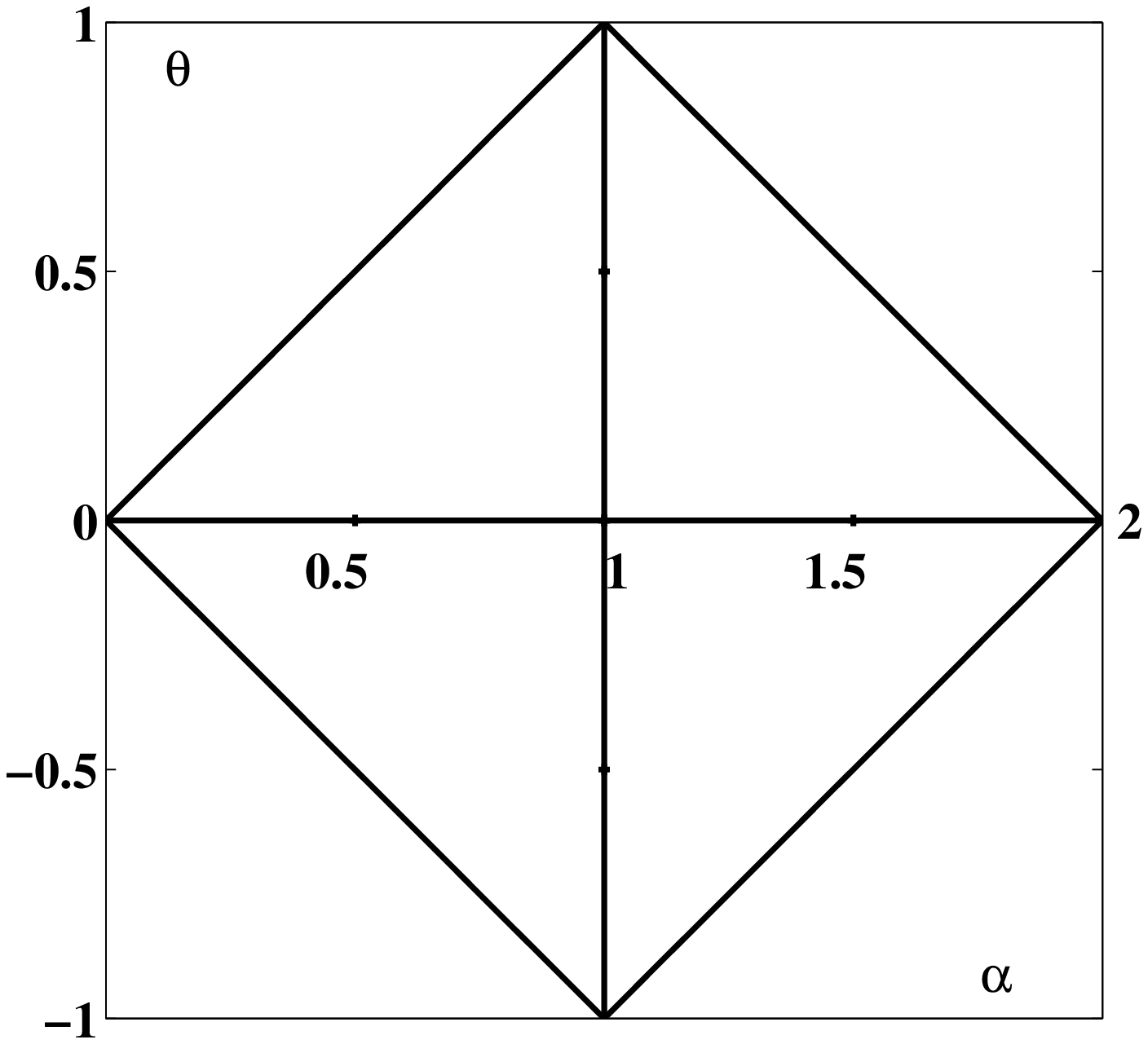}
\end{center}
\vskip -0.1truecm
\cen{{\bf Fig. 1} The Feller-Takayasu diamond}

\section*{Appendix C: The Mittag-Leffler   function}

The Mittag-Leffler function with parameter  $\beta$
is defined as  
$$ E_\beta (z) :=
    \sum_{n=0}^{\infty}\,
   {z^{n}\over\Gamma(\beta\,n+1)}\,, \q \beta >0\,, \q z \in \CC\,.
   \eqno(C.1)
 $$
It is an entire function of order $\beta $
and reduces for $\beta=1$ to $\exp (z)\,.$
For detailed information on  the functions of Mittag-Leffler type
the reader  may  consult \eg
\cite{Erdelyi_HTF,GorMai_CISM97,Kiryakova_94,MaiGor_JCAM00,
Podlubny_BOOK99,SKM_93} and references therein.
\vsp
Hereafter, we find it convenient to summarize
the  most relevant features  of the functions 
$$\Psi(t):= E_\beta(-t^\beta)\,, \q 0<\beta<1\,, \eqno(C.2) $$
$$\phi(t):=-  \frac{d}{dt}  E_\beta(-t^\beta) \,, \q 0< \beta <1\,,
\eqno(C.3)$$
 that turn out to be the  most relevant functions of Mittag-Leffler type for our purposes.
 Both of them reduce to the  exponential function $\exp (-t)$ in the limit
 as $\beta \to 1$.
\vsp   
We begin to quote their  expansions
in  power series of $t^\beta $ (convergent  for $t\ge 0$)
and  their asymptotic representations for  $t\to \infty $,
$$ \Psi(t)
   = {\ds \sum_{n=0}^{\infty}}\,
  (-1)^n {\ds {t^{\beta n}\over\Gamma(\beta\,n+1)}}
 \,\sim \,  {\ds {\sin \,(\beta \pi)\over \pi}}
  \,{\ds  {\Gamma(\beta)\over t^\beta}}\,,
     \eqno(C.4) $$
$$ \phi(t)
= {\ds {1\over t^{1-\beta}}}\, {\ds \sum_{n=0}^{\infty}}\,
  (-1)^n {\ds {t^{\beta n}\over\Gamma(\beta\,n+\beta )}}
 \, \sim \,  {\ds {\sin \,(\beta \pi)\over \pi}}
  \,{\ds  {\Gamma(\beta+1)\over t^{\beta+1}}}\,.
     \eqno(C.5) $$
\vfill\eject
\noindent
The Laplace transforms of $\Psi(t)$ and $\phi(t)$
can easily  be obtained by transforming the series (C.4), (C.5) term by term, respectively:
they read
$$
 \widetilde\Psi(s) = \frac {s^{\beta-1}}{1+s^\beta}\,,\quad
 \widetilde\phi(s) = \frac {1}{1+s^\beta}\,,\quad \Re s >0\,.\eqno(C.6)$$
For $0<\beta <1$ both  functions
 $\Psi(t)$, $\phi(t)$
keep   the complete monotonicity
 of the limiting exponential function of $\exp (-t)\,.$
Complete monotonicity of  a function
 $f(t)$   means,  for $n=0,1,2,\dots$, and  $t\ge 0$,
$ {\ds (-1)^n {d^n\over dt^n}\, f(t) \ge 0}$,
or equivalently, its representability as (real) Laplace transform
of a non-negative  
function or measure, see \eg \cite{Feller_71}.
\vsp
Recalling the theory of the Mittag-Leffler functions
of order less than 1,  we obtain
for $0<\beta <1$ the following representations, see \eg \cite{GorMai_CISM97},
$$ \Psi(t)  =
  {\ds{\sin \,(\beta \pi)\over \pi}\,
   \int_0^\infty \!
   { r^{\beta  -1}\, \e^{\,\ds -rt}\over
    r^{2\beta } + 2\, r^{\beta } \, \cos(\beta  \pi) +1}\, dr}\,,
  \q t \ge 0\,, \eqno(C.7)$$
$$  \phi(t) =
  {\ds {\sin \,(\beta \pi)\over \pi}\,
   \int_0^\infty \!
   { r^{\beta}\, \e^{\,\ds -rt} \over
    r^{2\beta } + 2\, r^{\beta}\,\cos(\beta \pi) +1}\, dr}\,,
 \q t\ge 0  \,. \eqno(C.8)     $$
In Figs 2 and 3 we exhibit plots of the
functions $\Psi(t)$ and $\phi(t)$, respectively
in logarithmic and linear scales.
 
\begin{center}
 \includegraphics[width=.49\textwidth]{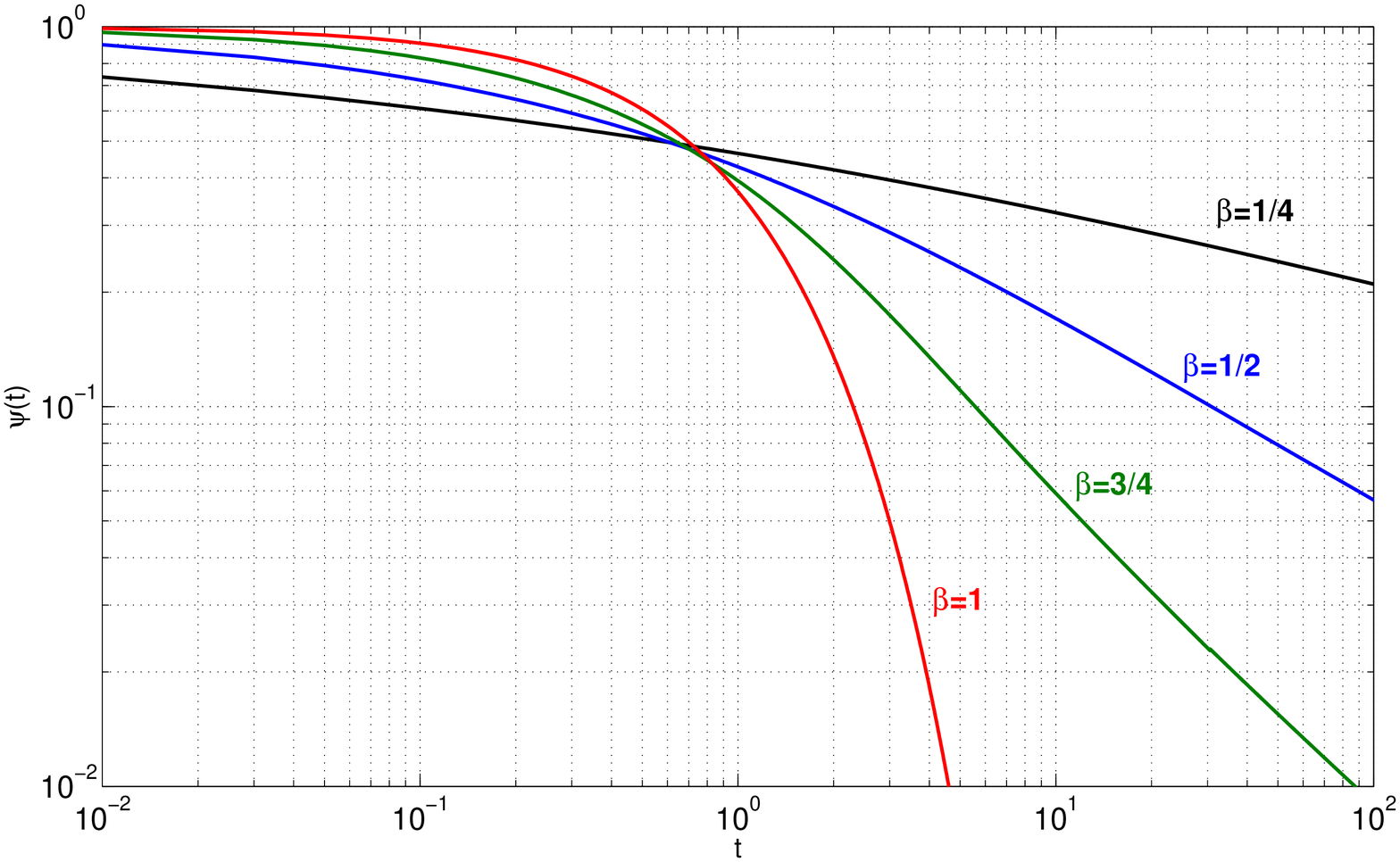}
\includegraphics[width=.49\textwidth]{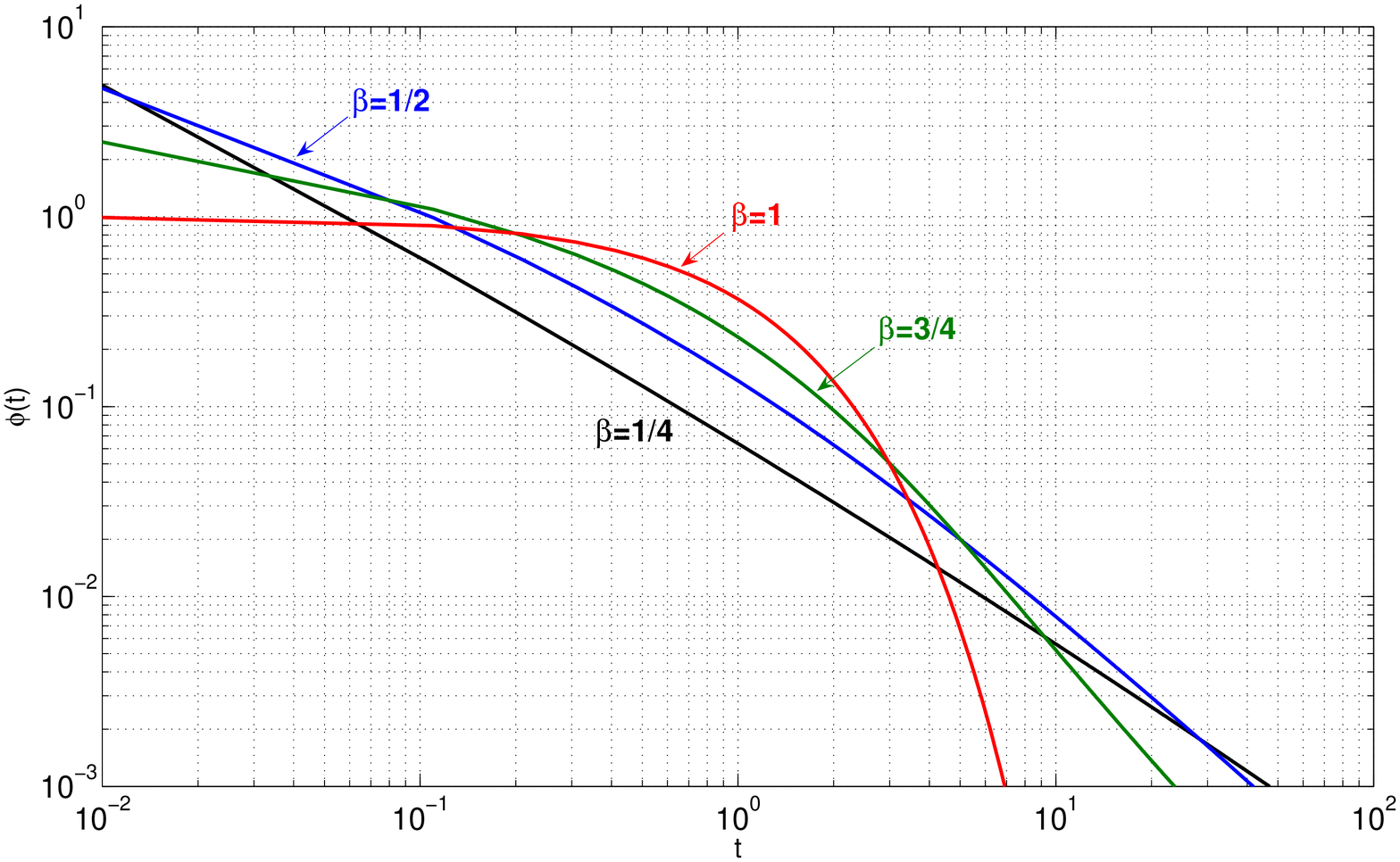}
\end{center}
\vskip -0.3truecm
\cen{{\bf Fig. 2} The functions $\Psi(t)$ (left) and
$\phi(t)$ (right) in logarithmic scales}
\vvs
\begin{center}
 \includegraphics[width=.49\textwidth]{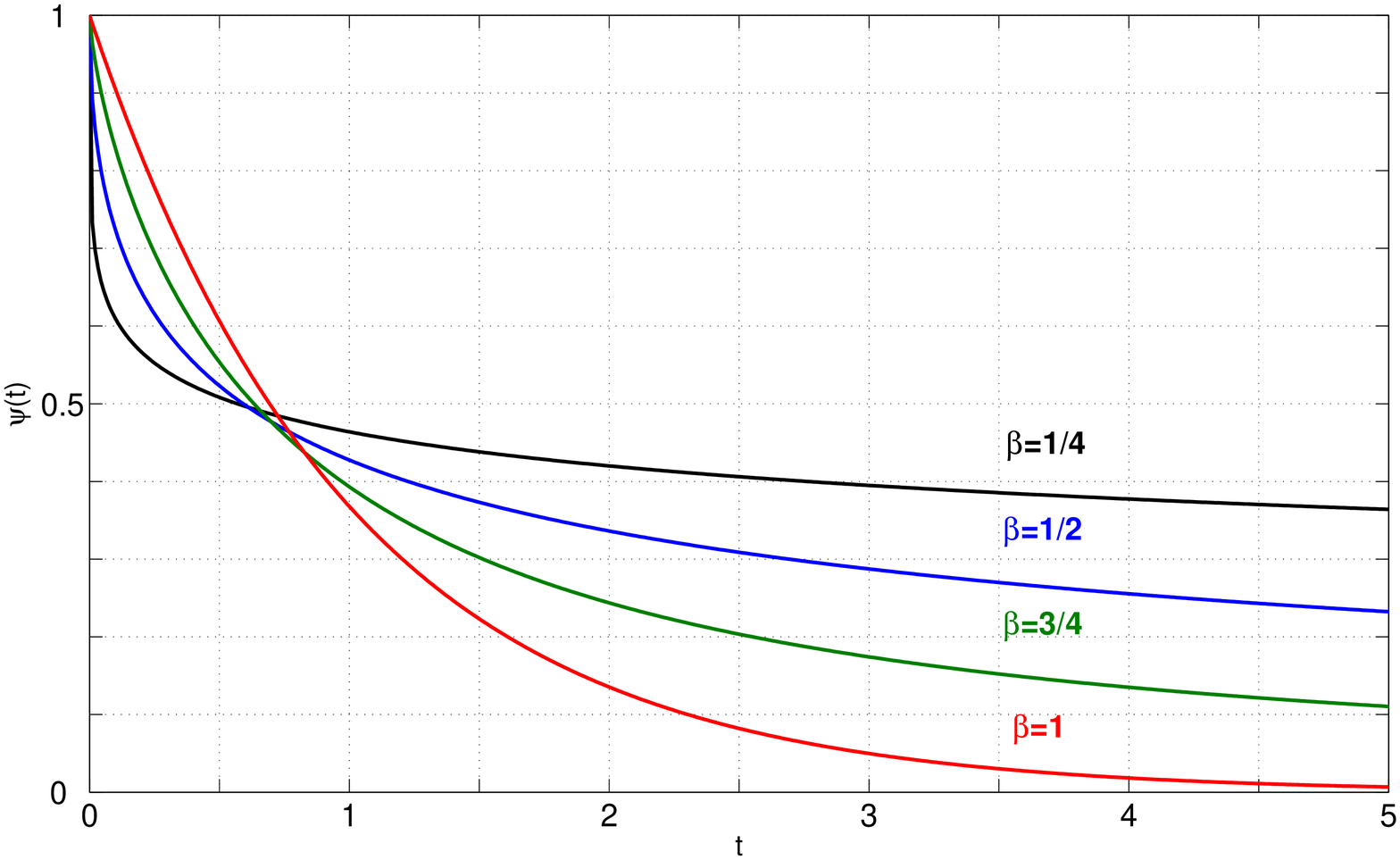}
\includegraphics[width=.49\textwidth]{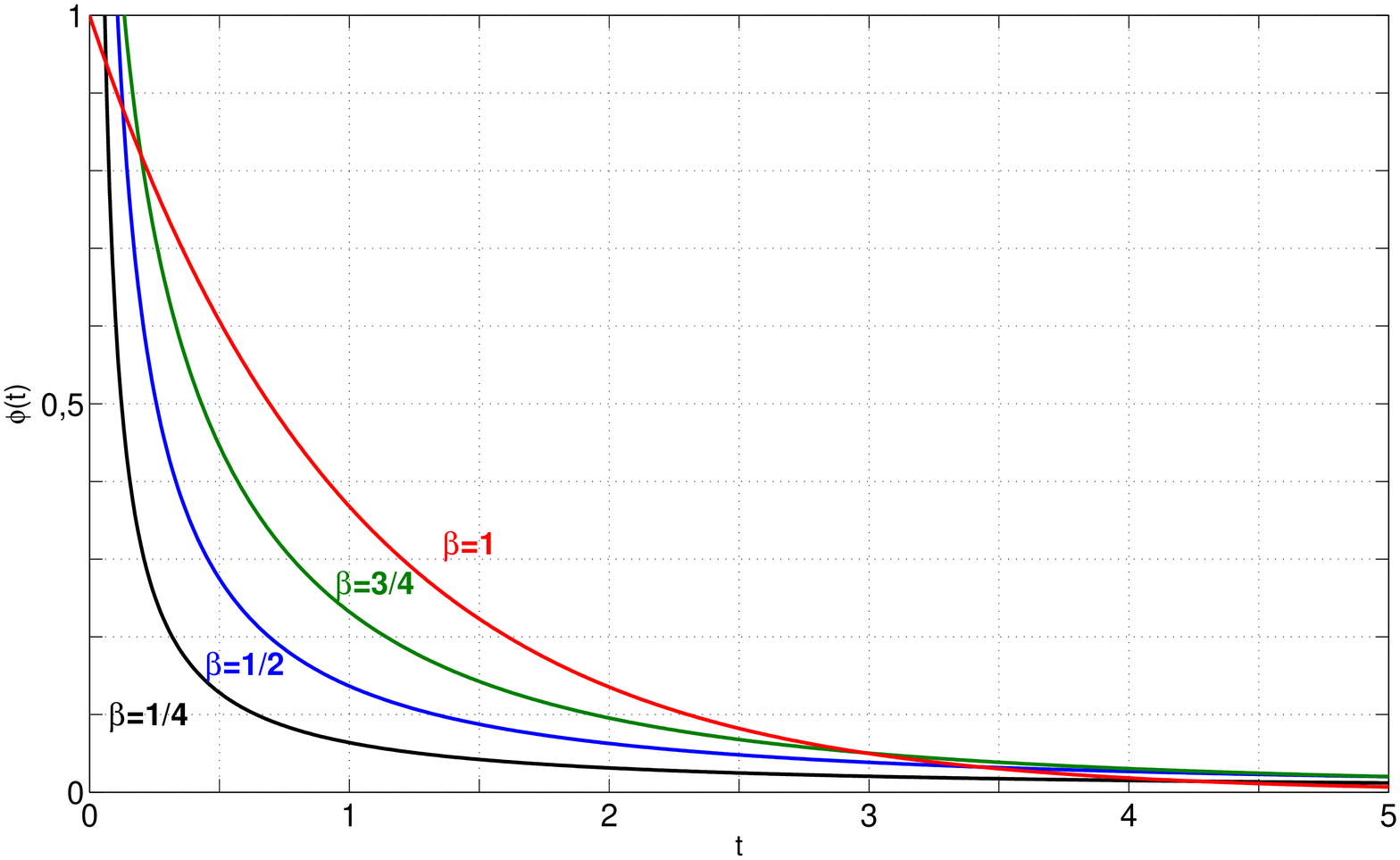}
\end{center}
\vskip -0.3truecm
\cen{{\bf Fig. 3} The functions $\Psi(t)$ (left) and
$\phi(t)$ (right) in linear scales}
\vfill\eject 

     \end{document}